\documentclass[fleqn,10pt]{wlscirep}
\usepackage[utf8]{inputenc}
\usepackage[T1]{fontenc}
\usepackage[nohyperlinks, nolist]{acronym}
\usepackage{bm}
\usepackage{siunitx}

\definecolor{dcol}{rgb}{0.7, 0.1, 0.1}

\definecolor{orcid}{rgb}{0.06, 0.4, 1.0}
\newcommand{\orcid}[2]{\href{https://orcid.org/#2}{ \textcolor{orcid}{#1}}}
\newcommand{\fdg}{\mbox{\ensuremath{.\!\!^\circ}}}%

\title{Analysis of the Breakthrough Listen signal of interest blc1 with a technosignature verification framework}

\author[1]{\orcid{Sofia Z. Sheikh}{0000-0001-7057-4999}}
\author[1, 2]{\orcid{Shane Smith}{0000-0003-4101-8234}}
\author[1, 3]{\orcid{Danny C. Price}{0000-0003-2783-1608}}
\author[4]{\orcid{David DeBoer}{0000-0003-3197-2294}}
\author[1]{\orcid{Brian C. Lacki}{0000-0003-1515-4857}}
\author[1]{\orcid{Daniel J. Czech}{0000-0002-8071-6011}}
\author[1, 5]{\orcid{Steve Croft}{0000-0003-4823-129X}}
\author[1]{\orcid{Vishal Gajjar}{0000-0002-8604-106X}}
\author[1,6]{\orcid{Howard Isaacson}{0000-0002-0531-1073}}
\author[1]{\orcid{Matt Lebofsky}{0000-0002-7042-7566}}
\author[1,4]{David H.E. MacMahon}
\author[1,5,7]{\orcid{Cherry Ng}{0000-0002-3616-5160}}
\author[8]{\orcid{Karen I. Perez}{0000-0002-6341-4548}}
\author[1, 5, 9]{\orcid{Andrew P.V. Siemion}{0000-0003-2828-7720}}
\author[1,10]{Claire Isabel Webb}
\author[11, 12]{\orcid{Andrew Zic}{0000-0002-9583-2947}}
\author[13]{Jamie Drew}
\author[13]{S. Pete Worden}

\affil[1]{Department of Astronomy, University of California Berkeley, Berkeley, CA, USA}
\affil[2]{Department of Physics, Hillsdale College, Hillsdale, MI, USA}
\affil[3]{International Centre for Radio Astronomy Research, Curtin University, Bentley, WA, Australia}
\affil[4]{Radio Astronomy Laboratory, University of California Berkeley, Berkeley, CA, USA}
\affil[5]{SETI Institute, Mountain View, California}
\affil[6]{Centre for Astrophysics, University of Southern Queensland, Toowoomba, Queensland, Australia}
\affil[7]{Dunlap Institute for Astronomy and Astrophysics, University of Toronto, Toronto, Ontario, Canada}
\affil[8]{Department of Astronomy, Columbia University, New York, NY, USA}
\affil[9]{Department of Physics and Astronomy, University of Manchester, Manchester, UK}
\affil[10]{Berggruen Institute, Los Angeles, CA, USA}
\affil[11]{Department of Physics and Astronomy, and Research Centre in Astronomy, Astrophysics and Astrophotonics, Macquarie University, Macquarie Park, New South Wales, Australia}
\affil[12]{Australia Telescope National Facility, CSIRO Space and Astronomy, Epping, New South Wales, Australia}
\affil[13]{Breakthrough Initiatives, NASA Research Park, Moffett Field, CA, USA}

\begin{abstract}
The aim of the search for extraterrestrial intelligence (SETI) is to find technologically-capable life beyond Earth through their technosignatures. On 2019 April 29, the Breakthrough Listen SETI project observed Proxima Centauri with the Parkes `Murriyang' radio telescope. These data contained a narrowband signal with characteristics broadly consistent with a technosignature near 982 MHz (`blc1').  Here we present a procedure for the analysis of potential technosignatures, in the context of the ubiquity of human-generated radio interference, which we apply to blc1. Using this procedure, we find that blc1 is not an extraterrestrial technosignature, but rather an electronically-drifting intermodulation product of local, time-varying interferers aligned with the observing cadence. We find dozens of instances of radio interference with similar morphologies to blc1 at frequencies harmonically related to common clock oscillators. These complex intermodulation products highlight the necessity for detailed follow-up of any signal-of-interest using a procedure such as the one outlined in this work.
\end{abstract}
\begin{document}

\flushbottom
\maketitle

\thispagestyle{empty}
\section{Introduction}

From 2019 April 29 to May 4, the \ac{BL} project performed observations to place limits on the prevalence of radio technosignatures (non-human ``objects, substances, and/or patterns whose origins specifically require a [technological] agent'', by analogy with biosignatures \cite{des2008nasa}) in the direction of \ac{ProxCen}. \ac{ProxCen} is an astrobiologically fascinating target due to its proximity: it is the closest star to the Sun at 1.295\,pc \cite{van2007validation}; it is host to Proxima b, the closest-known exoplanet to the Earth \cite{anglada-escude2016terrestrial, mascareno_2020}, which lies in the traditional habitable zone of \ac{ProxCen}; and it has even featured as the target of a proposed in-situ search via Breakthrough Starshot \cite{worden2018philanthropic}. 

We employed the CSIRO Parkes ``Murriyang'' telescope with the \ac{UWL} \cite{hobbs2020UWL}, across $0.704$--$4.032$\,GHz, as part of the project ``P1018: Wide-band radio monitoring of space weather on Proxima Centauri'' (Primary Investigator A.Z,). These observations were part of an international multi-wavelength campaign to monitor \ac{ProxCen} for stellar flares \cite{zic2020flare}, collaboratively carried out using the \ac{BL} Parkes Data Recorder backend in shared-risk mode \cite{price2018blpdr, price2021blpdr}. In parallel with the P1018 flare search, the \ac{BL} team used the data to conduct a technosignature search of \ac{ProxCen}. Over these six days, \ac{ProxCen} was observed for a total of 26\,h 9\,min; data are available at \url{seti.berkeley.edu/blc1}.

\begin{figure}
    \centering
    \includegraphics[height=0.8\textheight]{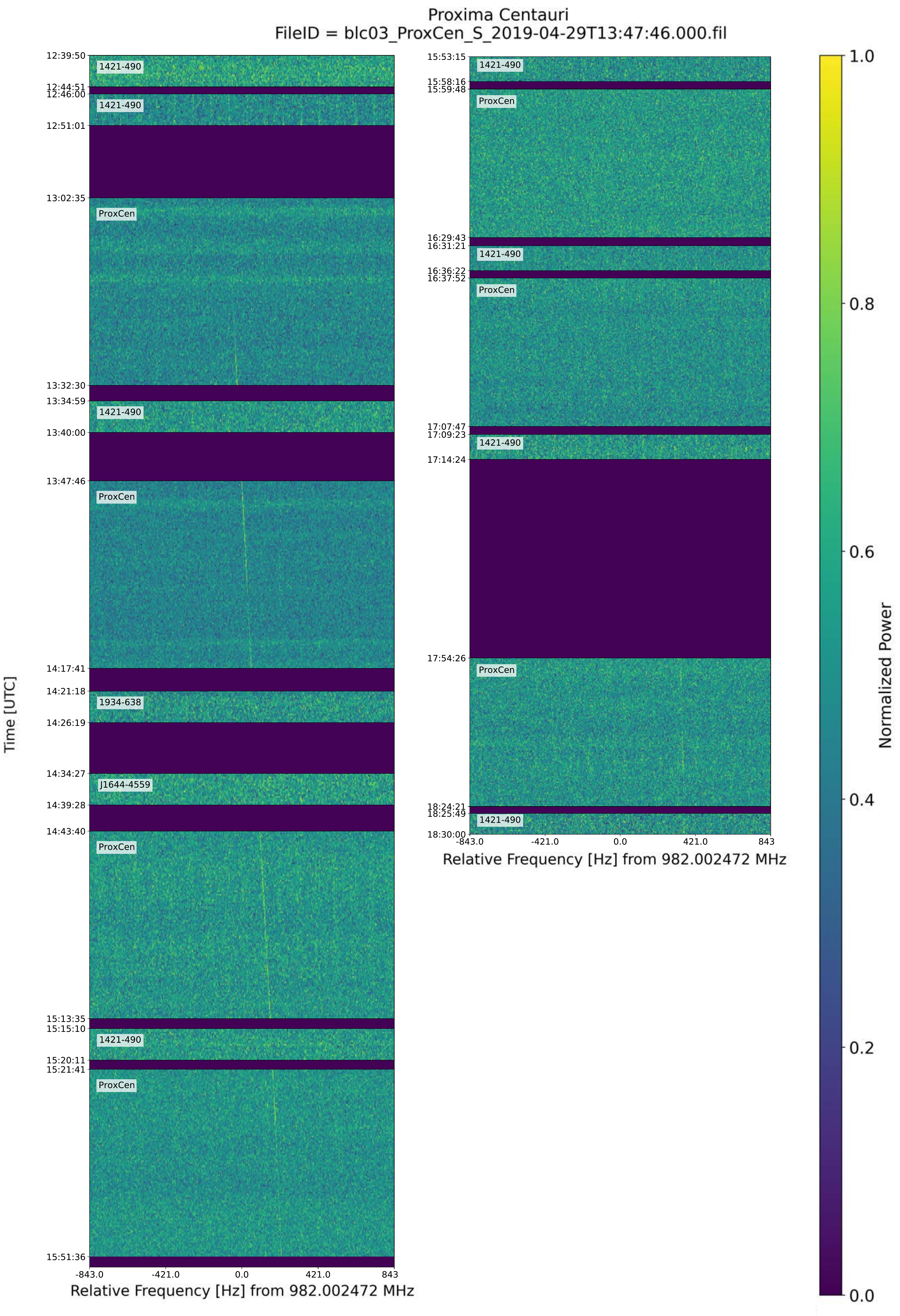}
    \caption{A waterfall plot (dynamic spectrum) around 982\,MHz depicting alternating observations of \ac{ProxCen} and off-source positions from 2019 April 29. The horizontal axis shows the relative frequency offset from a start frequency, the y-axis shows time progressing from top to bottom (and from left to right, continuing from the end of the first column to the beginning of the second), and the colorbar shows detected power (linearly normalized). blc1 is visible as the diagonal linear feature in ProxCen panels 1, 2, 3, 4, and 7. Data was not taken during purple panels, which are primarily telescope slews.}
    \label{fig:1}
\end{figure}

We searched for narrowband drifting signals in high-resolution dynamic spectra \cite{lebofsky2019} with $\sim 17$\,s subintegrations and $\sim 4$\,Hz frequency resolution. We ran the \texttt{turboSETI} (version 1.2.2) narrowband search algorithm---as described in a companion paper \cite{smith2020inprep}, and the pipeline returned a narrowband event that appeared in two consecutive observations of \ac{ProxCen} and did not appear in any reference ``off-source'' observations toward astronomical calibrator sources between the ``on-source'' \ac{ProxCen} observations, even upon visual inspection. This single event was named ``blc1'' as a shorthand for ``Breakthrough Listen candidate 1''. We find ``signal-of-interest'' to be a more appropriate categorization than ``candidate'' (see, for example, \cite{forgan2019rio} and Supplementary Discussion 1.1) but, as ``blc1'' is already in common usage, we will continue to use it here. blc1 is shown in Figure \ref{fig:1} and its properties are enumerated in Table \ref{tab:general_properties}.

\begin{table}[ht]
    \centering
    \begin{tabular}{|c|c|}
         \hline
         \textbf{Parameter} & \textbf{Value} \\
         \hline
         Detection Date & 29 April 2019 \\
         Time at First Detection & 13:17:35.232 UTC \\
         Time at Last Detection & 18:19:26.400 UTC \\
         Length of Persistence & 5.03 hours \\
         Signal Frequency at First Detection & 982.0024 MHz\\
         Signal Frequency at Last Detection & 982.0028 MHz\\
         Initial Drift Rate & 0.0326 Hz/s\\
         Average Signal-to-Noise & 17.956\\
         Signal Bandwidth & $<$3.81 Hz\\
         \hline
    \end{tabular}
    \caption{Basic characteristics of blc1. Average signal-to-noise is calculated as the average of the signal-to-noise ratios from the five 30-minute observations in which blc1 appeared.}
    \label{tab:general_properties}
\end{table}

blc1 is intriguing because:
\begin{enumerate}
    \item It is a $\sim$Hz-wide narrowband signal, which cannot be created by any known or foreseeable astrophysical system, only by technology \cite{tarter2001review}.
    \item It exhibits a non-zero drift rate, as expected for a transmitter that was not on the surface of the Earth.
    \item Its drift rate appears approximately linear in each 30 minute ``panel'' (one single-target observation from a cadence of observations that makes up a waterfall plot such as Figure \ref{fig:1}), but the drift rate changes smoothly over time, as expected for a transmitter in a rotational/orbital environment.
    \item It is absent in the off-source observations (see Section \ref{sec:initial}), as expected for a signal that is localized on the sky. 
    \item It persists over several hours, making it unlike other interferers from artificial satellites or aircraft that we have observed before.
\end{enumerate}

For these reasons, especially Point 4 (by which we have ruled out \texttt{turboSETI} outputs in prior searches, e.g., \cite{price2020breakthrough}) blc1 warranted an in-depth analysis beyond any event so far in the course of the \ac{BL} initiative.

\section{Initial investigation and parametrization of blc1}
\label{sec:initial}

We first ensured that the telescope and backend logs showed normal operation, mapped out the telescope pointings against the local site, and checked that data recording was functioning correctly. Upon consultation with the observatory and the Australian Communications and Media Authority (ACMA), there is no catalogued RFI at the Parkes Observatory at 982.002 MHz, nor registered transmitters at that frequency in Australia. The band near 982\,MHz more broadly is reserved for aircraft.

A signal that is truly localized on the sky should only be detected when the telescope is pointed at that location, if the off-source positions are suitably chosen with regard to the sidelobe pattern of the telescope. In order to ensure that the behaviour seen in the off-source observations is consistent with a localized emitter in the on-source, we must consider the cadence, off-source positions, and off-source luminosities, and perform a thorough search for highly-attenuated power on the off-source observations. 
The \ac{ProxCen} campaign employed a modified observing strategy that was different than the usual \ac{BL} procedures (\cite{enriquez2017turbo, price2020breakthrough}) in order to fulfill the project's primary goal: to observe and characterize stellar flares from \ac{ProxCen}. The cadence was non-standard, consisting of 30\,min integrations on \ac{ProxCen} that alternated with 5\,min observations of the calibrators. This cadence was chosen to maximize the on-source time. However, the asymmetry in the observing lengths (30\,min vs.\ 5\, min) causes an associated asymmetry in the expected \ac{S/N} from the same signal in a \texttt{turboSETI} analysis by a factor of $\sqrt{\frac{30}{5}}$.

In addition, the off-source positions were two quasars (and a single pulsar, observed once) selected from the ATNF calibrator database. Off-sources for \texttt{turboSETI} are usually chosen to be positions, usually stars, without a detectable radio flux, for consistency when evaluating apparent \ac{S/N}s between on- and off-sources. As these observations were being used for flare detection, the off-sources were chosen as quasars for frequent evaluation of the system temperature. The quasars, given their intrinsic radio emission, have a higher ``noise floor'' that we must contend with when searching for narrowband emission. From \cite{wright1990parkes}, we find a flux of $S(\nu) = 10$\,Jy at a frequency $\nu$ of 982\,MHz for both PKS\,1421-490 or PKS\,1934-638. Given our SEFD of $\sim$38 Jy from above, this leads to a total SEFD of 48\,Jy when observing on a calibrator.

Finally, the off-source positions were significantly further from \ac{ProxCen} than in a normal \ac{BL} observation. PKS\,1421-490 is $12\fdg57$ from \ac{ProxCen} and PKS\,1934-638 is $45\fdg82$. These distances led to the long slew times shown in Figure \ref{fig:1}, which can cause the temporal \ac{RFI} environment to change between the on and off-source observations. There may also be a large spatial difference in directional \ac{RFI}, especially when complicated by the telescope sidelobes. Similarly, the position of \ac{ProxCen} changed in sky coordinates by $~30^{\circ}$, primarily in azimuth, from first detection to final detection.

In this case, the first two effects would downweight the expected \ac{S/N} of blc1 in the off-source observations by a factor of $\sqrt{\frac{30}{5}} \times \frac{48}{38} \approx 3$, which should still be detectable in a de-drifted sum if the signal were present in the off-source. We took all nine off-source panels from Figure \ref{fig:1} and individually examined their waterfall plots, confirming that there was no visible signal in the first eight. The ninth and final off-source shows some power near 982.0028 MHz. 

Highly-attenuated signals in an off-source observation may only become apparent after removing the expected drift rate. To this end, we de-drifted each off-source panel to the nearest best-fit drift rates (see Supplementary Methods 2.1). The final off-source \textit{does} show a faint narrowband signal (S/N $<5$), but the best-fit drift rate for this signal is 0 Hz/s, not 0.14 Hz/s. While this could be interpreted as blc1 appearing in an off-source, we deemed it inconclusive given a) the lack of appearance in the previous off-source observations, b) the known presence of a 0 Hz/s interference comb in the spectrum (see Section \ref{ssec:comb}), c) the mismatch of the expected and measured drift rate, and d) the low \ac{S/N}.

Finally, in order to search for constant, highly-attenuated power across the off-sources, the panels can be individually summed across their 5 and 30 minute lengths to create integrated spectra, and those spectra can in turn be summed to produce an incoherent sum across every off-source observation. We saw no resulting signal (\ac{S/N} $ \gtrsim 5$) in the de-drifted spectra of any non-detection, nor any signal in the incoherent sum of the nine off-sources. This indicates that either a) the phenomenon was localized on the sky, either near \ac{ProxCen}'s location or near a source in the sidelobes of the telescope; or b) the phenomenon had a duty-cycle or brightness variation that was matching our chosen cadence particularly well. 

\subsection{Frequency Comb}
\label{ssec:comb}
Using an auto-correlation function in frequency, we find evidence for an \ac{RFI} frequency comb---a set of non-drifting signal regularly-spaced in frequency with stable amplitude---in the blc1 observation (Supplementary Figure 16). 
The comb is present in the on-sources and the off-sources, has a spacing of $\sim$80.1 Hz and is present throughout the entire 128 MHz sub-band, from 960--1087 MHz. This implies that the comb is not localized to the part of the spectrum around blc1, lowering the likelihood that they are related phenomena. To investigate this further, we searched for reappearances of the frequency comb in other \ac{UWL} observations (using the 7000 observation dataset described in Section \ref{ssec:archival}) and found that the comb appeared occasionally, throughout the year, without any associated signals at 982\,MHz. We do not detect the frequency comb in any of the November 2020 or April 2021 re-observations (Section \ref{ssec:reobs}.

We deem the comb to be independent of blc1 based on a) observing the comb over the week of \ac{ProxCen} observations and in unrelated campaigns throughout the year in both on and off-sources (indicative of \ac{RFI}) without an appearance of blc1; b) the lack of drift in the frequency comb; and c) a lack of correlation between the \ac{S/N} over time of the comb and the \ac{S/N} over time of the signal. However, the \ac{RFI} comb unfortunately complicates the measurement of parameters, especially the \ac{S/N}, in observations of similar signals to blc1 with heavy interference.

\subsection{Re-observations}
\label{ssec:reobs}

Around mid-November 2020, we began to re-observe \ac{ProxCen} during our scheduled observation time at Murriyang, for sessions when it was above the horizon. We re-observed \ac{ProxCen} on November 19, 26, and 30, for 3--4 hours on each occasion. These re-observations were performed with the same \ac{UWL} receiver, \ac{BL} backend, and outputted data product as the original observation of blc1. However, we chose a more standard cadence for the re-observations: 15 minutes on-source on \ac{ProxCen}, 15 minutes off-source, with the off-sources chosen to be nearby stars at different relative angles to the \ac{ProxCen}--ground vector. No signals of any drift rate were detected at $\nu_{\rm blc1} \pm 2$\,kHz in any of these three re-observations.

Two years after the original detection, from April 29--May 3 2021, we observed 12 hours each day to replicate the relevant portion of the initial \ac{ProxCen} session in 2019. These re-observations were performed with the same \ac{UWL} receiver, \ac{BL} backend, outputted data product, local time, and cadence as the original observation of blc1. No signals of any drift rate were detected at $\nu_{\rm blc1} \pm 2$\,kHz in any of these re-observations.

\section{Constraining physical and electronic drift rates}

Signals drifting in frequency are generated in one of two ways: (1) a transmitter accelerating relative to the observer and (2) an electronically varying (deliberately or not) transmitter. Both means may be present at any given time. 

Strong ground-based \ac{RFI} entering into a distant sidelobe may remain detectable even when the telescope is pointed at different parts of the sky. The sidelobe may downweight the signal by $10^5$ (50\,dB) or more \cite{hobbs2020UWL}, but it may still be detected with higher \ac{S/N} than an intrinsically weak and/or distant signal coming through the main lobe with as little as one millionth of the flux density. For satellites, the main lobe and near-in sidelobes subtend a very small area so ``direct hits'' from satellites are rare but can saturate the receiver when they do occur. 

\subsection{Accelerational Drifts}

For this analysis, we assume that the signal from a transmitter near 982\,MHz is leaking in through either a distant antenna sidelobe or the main beam. We seek to explain both the drift morphology of the signal and the prevalence over a sufficient length of time. 

With the frequency fixed, we looked at drift rate characteristics of transmitters undergoing various ``normal'' motions. The drift is proportional to the relative line-of-sight acceleration between the receiver and transmitter, which can be produced via a change of speed or relative direction. The drift due to just telescope motion as it tracks an object is $<10^{-4}\dot{\nu}$, where $\dot{\nu}$ is the signal drift of blc1, and hence does not explain blc1's characteristics. 

\subsubsection{Ground-Based and Aerial Transmitters}

A transmitter on the ground may be stationary, or be on a moving platform in the vicinity of the telescope, such as a car, train or other vehicle.  In the air, the transmitter could be on a plane, helicopter, drone, balloon or other airborne object. To investigate these hypothetical transmitters, we looked at a range of plausible routes around the observatory (say for someone walking or biking) and for a range of velocities for vehicles along the nearby highway, as shown in Supplementary Figure 1, for a fixed frequency transmitter. Similarly, we explored a range of plausible airplane velocities and routes between various cities.

Supplementary Figure 1 shows the primary issue with all of the local signals described above: that speeds fast enough to exhibit the right order of drift do not persist long enough; conversely, for slow enough speeds to persist, the drifts are too low. None of the drift characteristics from local signals match the measurement and it is extremely difficult to construct a continual motion path that could persist as exhibited by the measured signal, even by varying the speed along the route. 

\subsubsection{Satellite Transmitters}

Next we considered artificial satellites orbiting the earth, which are broadly grouped as (1) low-earth orbit (LEO, $\sim$90 minute orbits), (2) medium-earth orbit (MEO, $\sim$12 hour orbits), and  (3) geosynchronous orbit (GEO, $\sim$24 hour orbits). The short period of LEOs implies that they cannot be responsible for the signal regardless of their drift characteristics, so we did not include LEO satellites in our analysis. Conversely, we do include GEO satellites; GEO is often referred to as ``geostationary'', implying true zero drift, but even stable GEO satellites generally meander in a figure-8 motion. There are a small number of satellites in other orbits as well, which are encompassed by the following analysis. 

\begin{figure}
    \centering
    \includegraphics[width=\textwidth]{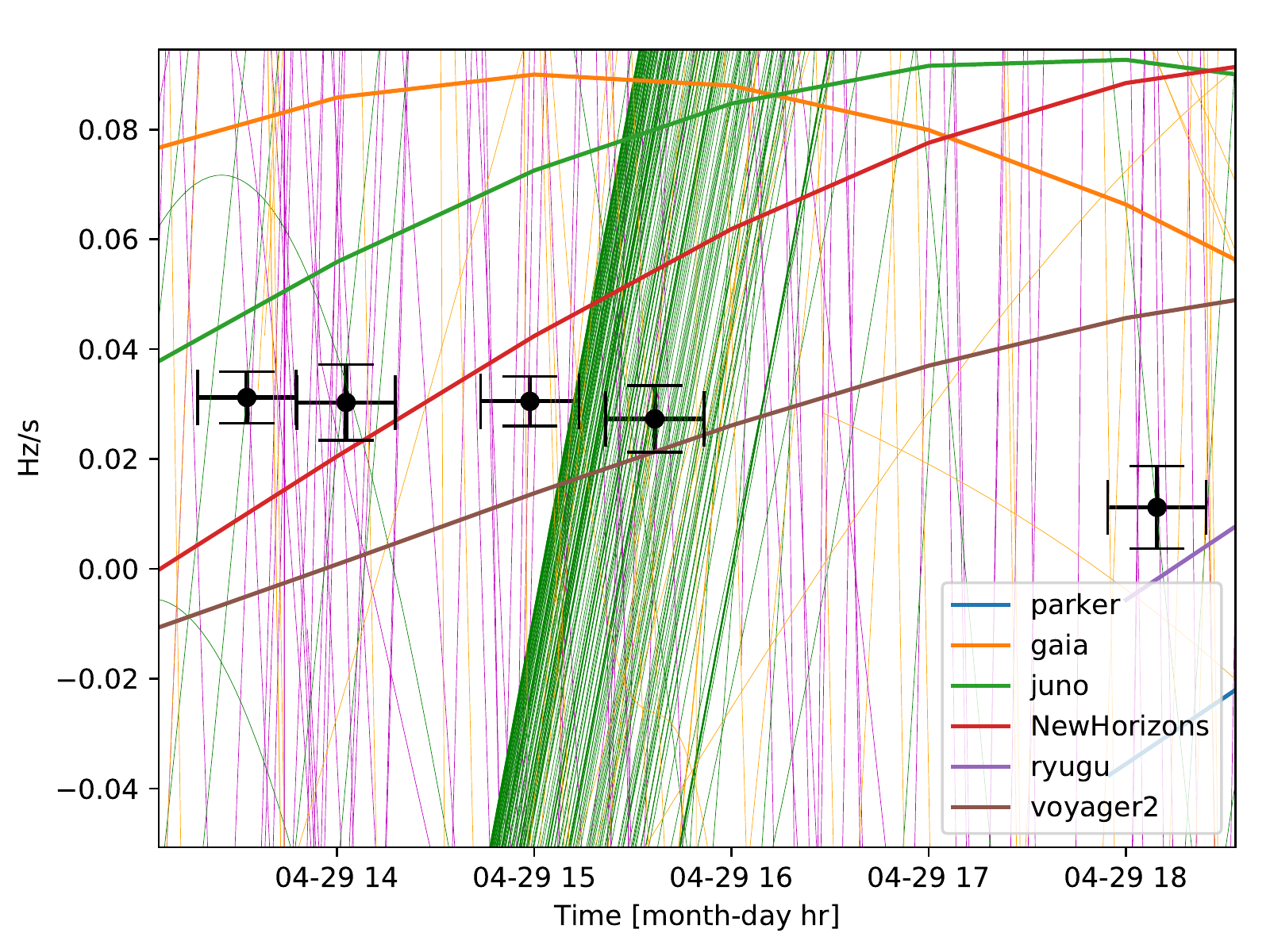}
    \caption{Drift characteristics of non-LEO satellites and deep space probes above the horizon during the event time. The thin green lines are GEO, the magenta are MEO, and the orange are other non-LEO satellites. The thick lines are deep space probes: Parker Solar Probe (blue), Gaia (orange), Juno (green), New Horizons (red), Ryugu (purple) and Voyager 2 (brown).  The black data points show the blc1 data, where the x-axis bars are the time of observation and the y-axis bars are 1$\sigma$ error bars. The deep-space probes are closer to the correct order of blc1’s drift due to their unique status as near-sidereal sources, but none can explain blc1.}
    \label{fig:2}
\end{figure}

To investigate these satellites, we downloaded orbital element parameters for all known, active, non-LEO satellites for the date in question (\url{https://www.space-track.org}). All satellites above the horizon are plotted as the faint lines in Figure \ref{fig:2}. Even for geosynchronous satellites (green lines), the drift characteristics vary too rapidly to explain the signal characteristics.

\subsubsection{Transmitters on Deep-Space Probes}

Deep space probes are exploratory spacecraft that do not orbit the Earth, and are the most celestially-stationary sources produced by humans.  Though their radio emission can be strong relative to celestial sources, their distance generally makes them weak enough that they are not readily detectable unless they happen to be in or near the main beam. To investigate space probes, we obtained positional information from NASA Horizons \url{https://ssd.jpl.nasa.gov/horizons.cgi}. Figure \ref{fig:2} shows the drift of space probes above the horizon, though none of them coincide with the main beam of the telescope. Their drifts and positions are inconsistent with blc1.

We also considered---and dismissed---transmitters on asteroids and reflections from Earth-bound radio transmitters off of asteroids in the primary beam as a potential source of blc1 (see Supplementary Methods 2.2).

\subsection{Electronic Drifts}

Finally, we considered electronically-varying signal generators. A modern signal generator can be programmed to produce any drift rate; there are effectively no constraints, and thus we cannot use this to inform our analysis. However, we can constrain other causes of electronic drift related to the temperature, aging and voltage of the oscillator---all oscillators exhibit some drift with these parameters. 

This type of \ac{RFI} is prevalent, but typically the drifts are too low (very good regulation) or too high and wandering (as in more typical commodity devices) for them to be mistaken for an object moving sidereally.  They are also almost always seen in both the off-source pointings as well as the on-source pointing.  However, given their prevalence, occasionally a device could potentially exhibit the expected characteristics of a sidereal source, which could be difficult to ascribe to \ac{RFI}.

\subsection{Astronomically-Expected Drifts}

We can also compare the drift rate magnitude and morphology to the sidereal drift in the direction of \ac{ProxCen}, as well as known accelerations and orbital periods from the planets in the \ac{ProxCen} system. Defiance of expectations for drift rates in a target system does not invalidate a signal, but a match provides additional evidence in favour of an interstellar origin. In an effort to be easily detected and discernable as ``stationary'', a distant transmitter in the direction of the antenna pointing could electronically vary their transmission frequency to compensate for their motion relative to the Sun, the Earth, or the barycenter of the solar system. For the timescales relevant to our observation, the Earth's rotation is the primary contributor to the expected drift rate for \ac{ProxCen}, and is similar for all other sidereal sources in the beam (see Supplementary Discussion 1.2). Supplementary Figure 2 shows the relevant geometry of the Earth-Sun-ProxCen positions and the corresponding residual drift. We find that the drift rate of blc1 is not consistent with the barycentric motion expected from the direction of \ac{ProxCen}, but is consistent with the order-of-magnitude drift that could be produced in the system, based on the orbital and rotational motions of its planets (see Supplementary Discussion 1.3). 

\section{Searching for other instances of blc1}

In parallel with the drift rate analysis, we performed a search for reappearances of the signal-of-interest on other days and at other frequencies. 

\subsection{Other Murriyang signals near 982 MHz}
\label{ssec:archival}

We searched for signals with the same frequency and same drift as blc1 from both a) the week-long \ac{ProxCen} campaign and b) every archival observation from standard \ac{BL} Murriyang \ac{UWL} observing of other stellar targets.

In order to find all appearances of blc1, even those that were too faint or masked by \ac{RFI} to be flagged by \texttt{turboSETI}, we produced output plots for visual inspection from every \ac{ProxCen} on- and off-source observation from April 29 to May 4. We restricted the plots to $982.002 - 982.004$\,MHz to begin; a few plots were extended up or down by 1\,kHz if there appeared to be interesting behavior near the upper or lower bounds. 

We created and searched two kinds of output plots: dynamic time-frequency spectra ``waterfall'' plots (see Figure \ref{fig:1}) and ``butterfly'' plots, which display the power at each drift rate-frequency pair after use of a de-drifting algorithm (see Supplementary Methods 2.1). Examples of these output plots are shown in Supplementary Figure 17. 

Through analysis of these diagnostic plots, we identified four occurrences of blc1-like signals during the \ac{ProxCen} observations which, through their low \ac{S/N}s, had failed to reach \texttt{turboSETI}'s detection threshold. An example of one of these similar signals is shown in Figure \ref{fig:3}, while the rest are displayed in Supplementary Figures 3--5. 

We performed the same analysis with every archival observation that \ac{BL} had taken with the Murriyang \ac{UWL} receiver. In total, this consisted of about 7000 observations from 2019--2020---primarily nearby stars in the Hipparcos catalog \cite{perryman1997hipparcos,isaacson17}, but also pulsars and quasars used for calibration, and, of course, \ac{ProxCen}. Most of these files had a standard duration of 5\,min, as opposed to the 30\,min observations from the \ac{ProxCen} campaign. In many of these observations, we observed the same zero-drift, frequency comb as detected in the original blc1 observation (see Section \ref{sec:initial}), giving more evidence that this comb is unrelated \ac{RFI}. 

We found 15 similar-looking features in these non-\ac{ProxCen} \ac{UWL} observations. Upon visual inspection of the full cadences surrounding these features, 14 of those are different from the blc1-like signals in morphology, length, drift behaviour, and/or signal strength over time. However, one of the 15 features is clearly \ac{RFI} due to its persistence across both on and off-source observations, appeared 4 days \textit{prior} to blc1, and looks to potentially be a member of the set including blc1 and the similar signals from the ProxCen campaign; this signal is shown in Supplementary Figure 6.  

\begin{figure}
    \centering
    \includegraphics[height=0.8\textheight]{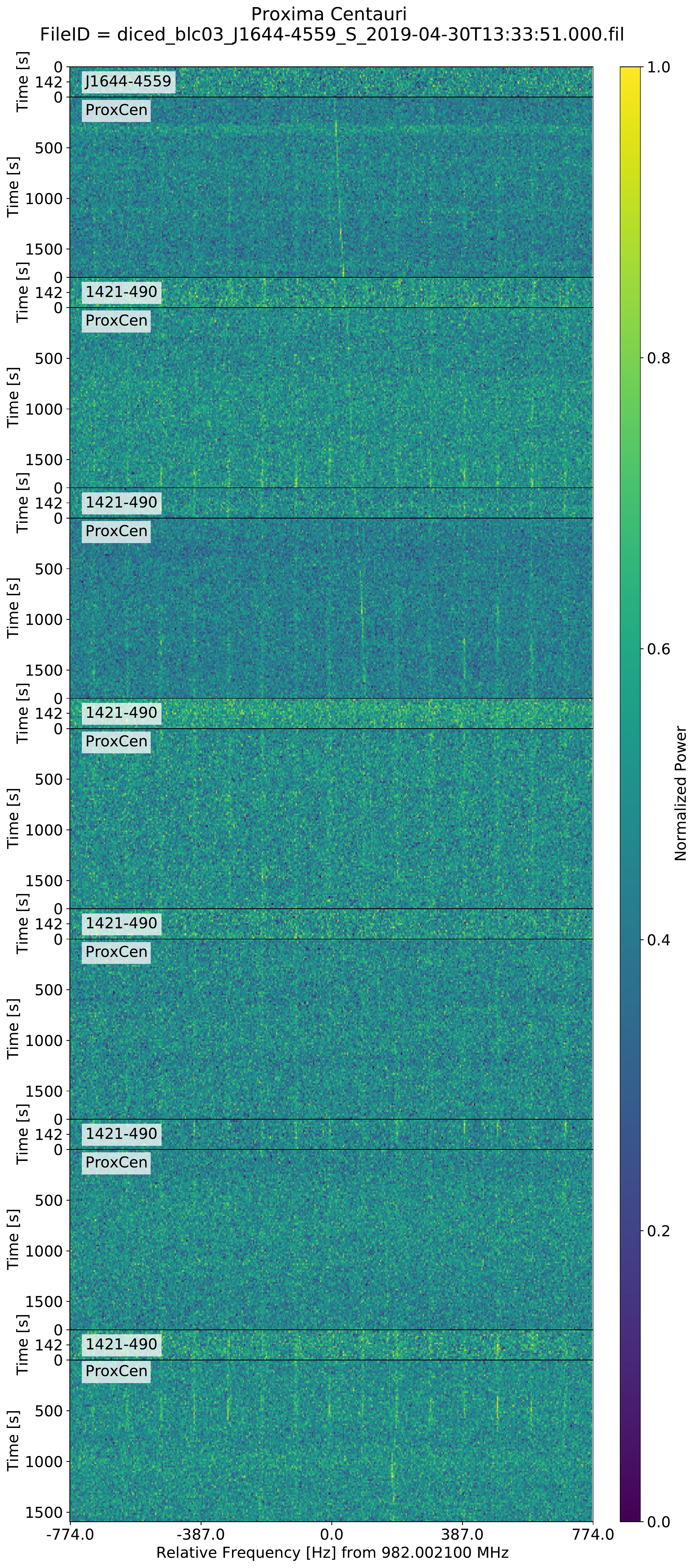}
    \caption{A waterfall plot around 982\,MHz depicting a similar signal to blc1 that occurred on 2019 April 30. The signal is the bright sloping feature, but is clearly overlaid on a non-drifting, comb-like signal (see Section \ref{ssec:comb}). Both the signal and the comb underneath appear to be \ac{RFI} as they appear in the off-source observations. In Supplementary Table 1, we derived the \ac{S/N}, drift rate, and start frequency of each panel of the blc1 observation; we derived the properties of this signal in the same manner (see Supplementary Methods 2.1). The drift rate (median 0.021 Hz/s), \ac{S/N} (median 6.9), and frequency range (982.0021--982.0023 MHz), are consistent with blc1, and this signal is also unresolved.}
    \label{fig:3}
\end{figure}

All five of these 982 MHz signals from different days are fainter than blc1; three of them conclusively appear in the off-sources, while two of them are inconclusive.

\subsection{Different Frequencies}

Human-made communication technologies often use multiple simultaneous frequency channels to send information for improved redundancy and bitrate. It is possible that \ac{ETI} would do the same; it is also possible that the appearance of a blc1 twin at another frequency, if clearly \ac{RFI}, would allow us to determine that blc1 is also \ac{RFI}. 

To find similar signals to blc1 at different frequencies, we calculated the frequency-normalized drift rate ($\dot{\nu}_{\rm normalized} = \frac{\dot{\nu}}{\nu}$) that would indicate a signal which was drifting proportionally to blc1 in the first observation. This proportional drift is expected for multi-frequency transmitters in the same accelerational environment, but also in multi-frequency transmitters that are electronically-drifting. 

We then searched the catalog of narrowband hits created with \texttt{turboSETI} for signals that a) appeared at the same time as blc1 and b) were drifting proportionally in the first panel, plus or minus the drift rate error proxy as given in the first row of Supplementary Table 1. This search returned 112 hits: blc1 itself, and 111 signals that \texttt{turboSETI} had identified as hits but then rejected as \ac{RFI} at an early stage in the pipeline due to their appearance in off-source observations. These hits were plotted in the context of all of the subsequent panels of the \ac{ProxCen} observation on 2019 April 29. 

We visually inspected the 111 matches, looking for signals that had the same morphology as blc1 beyond the drift rate in the first panel. To identify a signal with the same morphology, we looked for monotonicity, a shallowing of the slope over time, a vertical length that spanned multiple panels, and the absence of complex additional features (e.g. sinusoidal behaviour). We found that 36 of the 111 \texttt{turboSETI} matches (32\%) were blc1 ``lookalikes'': signals with strikingly similar morphology to blc1. A subset of these lookalikes are shown in Supplementary Figure 7, while some examples of the ``non-lookalikes'' are shown in Supplementary Figure 8. 

The lookalikes have a range of variabilities over time, which seem to indicate multiple transmitters producing unusually consistent drifts. We then performed the same search as before but with negative drift rates. We found 310 hits from \ac{RFI}, of which 27, upon visual inspection, were found to be \textit{mirrored} lookalikes: with exactly the same drift structure over time as blc1 but flipped in morphology across the frequency axis. A selection of these signals are shown in Supplementary Figure 9.  

We can conclusively state that all lookalike and mirrored lookalike signals are \ac{RFI} due to their appearance in off-source observations. 

\subsection{Characterizing the blc1 lookalike population}

\begin{figure}
    \centering
    \includegraphics[width=\textwidth]{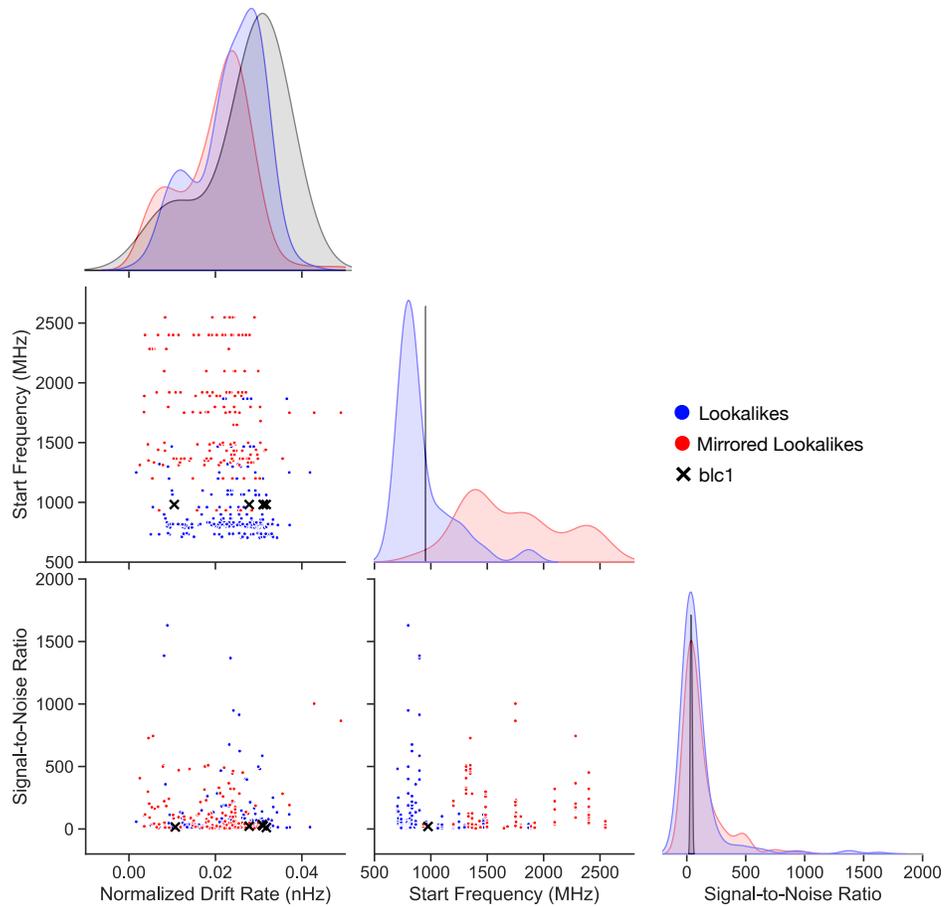}
    \caption{A corner plot showing the population of lookalikes and mirrored lookalikes compared to blc1 in start frequency, normalized drift rate, and signal-to-noise ratio. For the mirrored lookalikes, which have negative normalized drift rates, we took the absolute value for comparison. The height of the kernel density estimations are not to scale, as the population of blc1 points is so much smaller than the other two populations that it would otherwise not be visible on these axes. The blc1 signal (black x symbols and grey shading) is consistent with the signal-to-noise and normalized drift rate distributions (blue points and shading for lookalikes, red points and shading for mirrored lookalikes) and, although slightly higher in frequency than the peak of the kernel density estimations for the lookalike population, is still consistent with that distribution.}
    \label{fig:4}
\end{figure}

The presence of this population of both positively- and negatively-sloped blc1 lookalikes preliminarily suggests that all lookalikes (including blc1) share a common origin. We can further assess this claim by examining the similarity of the parametric distributions of the lookalike signals and blc1. We find that blc1 is consistent with the lookalike population in absolute drift rate, frequency, and \ac{S/N} (Figure \ref{fig:4}).

\subsection{Determining the origin of the blc1 lookalike population}

We can further strengthen the claim that the lookalike population shares a common origin if we can identify a frequency-shifting instrumental or electronic effect in the data. One potential source of that effect is instrumental harmonic distortion, which can produce replicas of an original frequency $f$ signal at $2f$, $3f$, ..., etc. Another potential source is intermodulation distortion, a superset to harmonic distortion, which can produce a near-arbitrarily complex sequence of replicas which are integer multiples of the sums and differences of two or more original signals.

\subsubsection{Harmonic Analysis}
\label{sec:harmonic}

We investigated whether any of the positive-drift lookalikes could be linked via a harmonic sequence. We generated the first 20 harmonics of a range of fundamental frequencies starting outside the bandpass at 100 MHz and progressing to 1000 MHz in 1 kHz intervals. We defined a potential ``harmonic sequence'' within the data as a set of two or more lookalikes (blc1 included) associated by the same fundamental frequency, within 1 kHz of their theoretical values. blc1 was not consistent with being in a harmonic sequence with any of the observed lookalikes. However, a harmonic sequence does interlinking a set of other lookalikes (Supplementary Table 2). This harmonic sequence contained frequencies in the form  $n + 0.1m + 0.099$ MHz, where $m$ and $n$ are integers---because of the constant term, we refer to this set as ``x.y99''. 

\subsubsection{Intermodulation Analysis}

Some lookalikes showed additional frequency structure that was not present in blc1. Two sets of lookalikes, which we will refer to as ``Triple Feature'' (TF) and ``Single Feature'' (SF) were distinguishable from their morphology alone. An example from the TF set is shown in Supplementary Figure 10. TF and SF contained both positive lookalikes and mirrored lookalikes. TF had spacings which were integer multiples of 133.33\,MHz. SF had a more complicated relationship of spacings: primarily integer multiples of 15\,MHz, but with an additional appearance of 128\,MHz. In all cases within both sets, these frequencies are consistent with common clock oscillator frequencies used in digital electronics, with matches within 1--1000\,Hz of the expected value. 

The three initially-identified sets---TF, SF, and the harmonic ``x.y99'' set (Section \ref{sec:harmonic})---each have a transition region where the morphology of the signal flips, with positive lookalikes on one side and mirrored lookalikes on the other (Supplementary Figure 11). If intermodulation effects were present, we predicted that we should detect a strong, zero-drift interferer at a frequency within the transition region, whose position is dictated by clock oscillator frequencies previously identified in the set. 

For TF, we find a strong interferer at the predicted central frequency of 1400\,MHz, consistent to a single channel ($\pm$ 4\,Hz). For SF, one central frequency consistent with the 30\,MHz spacing is 1200\,MHz, where we also see a strong interferer consistent at the Hz level. For x.y99, we predict a central frequency of 1332\,MHz, but see only an extremely faint interferer; we do see strong signals at 1330.0000\,MHz, 1331.2000\,MHz, and 1332.1805\,MHz. This inconsistency could be caused by additional transposition, from an additional oscillator, present within the set, implying that one of the three frequencies listed prior is actually the responsible interferer. We conclude that these intentional-seeming separations seem likely to have been produced by the intermodulation of at least one clock oscillator with a strong interferer. 

\begin{figure}
    \centering
    \includegraphics[width=\textwidth]{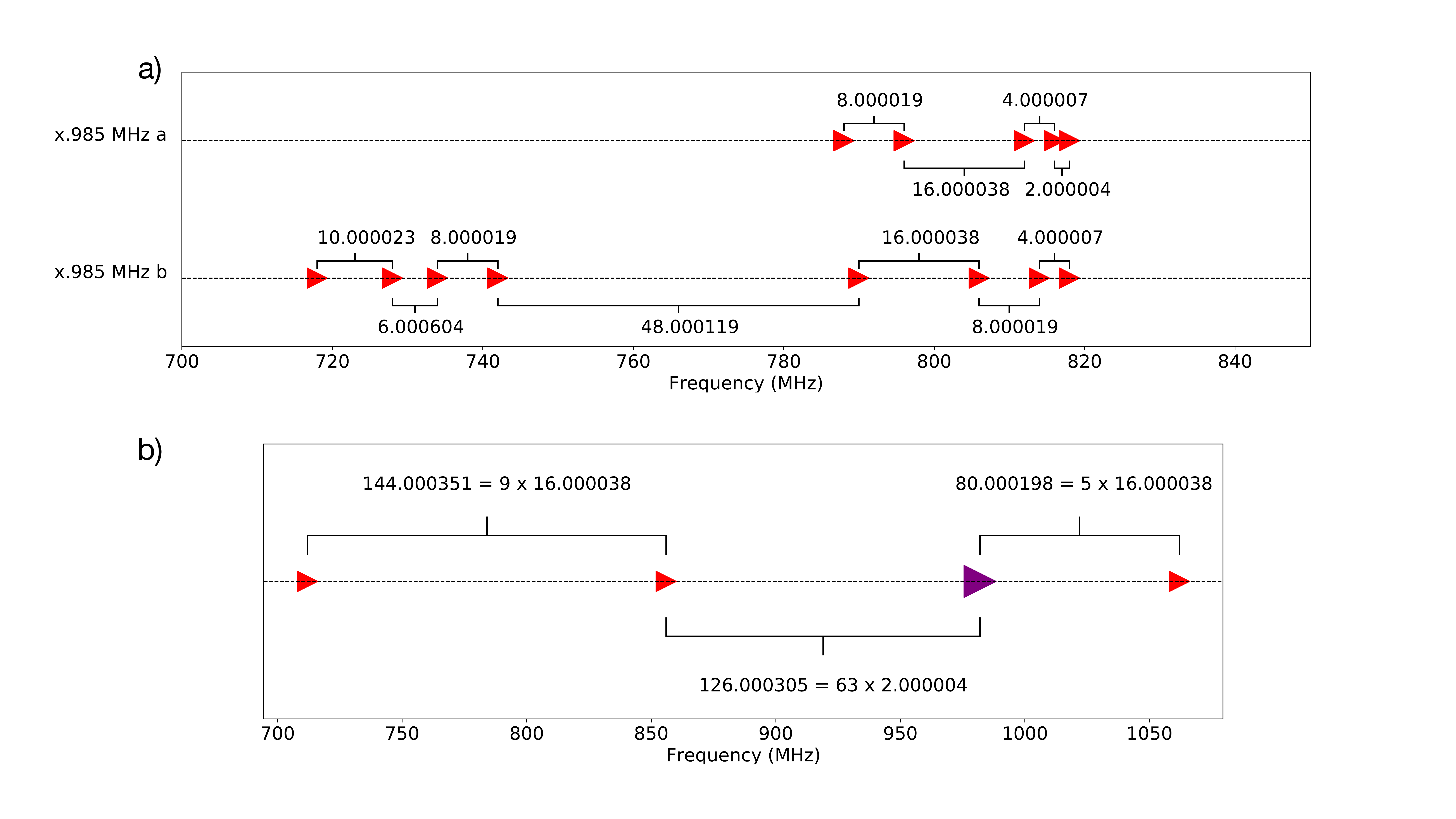}
    \caption{A visualization of 16 signals from the lookalike set and blc1, divided into three sets. Top Panel a): Two sets of positive lookalikes found within the data, originally identified by their position at 985\,kHz above an integer MHz in each appearance. The full spacings, to Hz precision, are shown to demonstrate the set's consistency with mixing from a 2.000004\,MHz clock oscillator. Recall that the spacings have an inherent, multiplicative $\pm 4$Hz error from the discrete frequency resolution. One spacing we see in both sets, apparently resulting from this same oscillator, is 16.000038\,MHz. The two sets are both being affected by the same oscillator, but they are not consistent with \textit{each other} to a multiple of 2.000004\,MHz, illustrating another complexity within the dataset. Bottom Panel b): blc1 (purple) shown in sequence with three additional lookalikes (red) which are consistent with integer multiples of 2.000004\,MHz, including exactly 16.000038\,MHz: a set of clock-oscillator induced spacings that is perfectly consistent with the spacings uncovered in the sets in a).}
    \label{fig:5}
\end{figure}

We found evidence for additional clock oscillator frequencies affecting the lookalike population, including the sets displayed in Figure \ref{fig:5}. In the top panel of Figure \ref{fig:5}, we searched for patterns with integer kHz offsets (as in Section \ref{sec:harmonic}) and uncovered two individual sequences with spacings consistent with a 2.000004\,MHz clock oscillator. Our data have a resolution of $\sim 4$\,Hz, so we expect that the last digit has an error of $\pm 4$, which will propagate in the integer multiplication of spacings. In Figure \ref{fig:5} we display the extra error at the Hz-level to illustrate that these spacings are clearly the product of the same oscillator: not only are the general spacings consistent with powers of 2.000004\,MHz, but the \textit{errors} on those spacings are consistent with propagating errors of order $\pm 4$ Hz, e.g., 16.000038\,MHz. In the bottom panel of Figure \ref{fig:5}, we see these exact spacings relating blc1 (within propagated error, of order 100\,Hz) with three lookalikes at $\sim 712$\,MHz, $\sim 856$\,MHz, and $\sim 1062$\,MHz. 

This numerical analysis indicates that blc1 is an intermodulation product being produced by a $\sim 2$\,MHz clock oscillator being mixed with some other zero-drift \ac{RFI} elsewhere in the band. We also find power at the blc1-companion lookalike frequencies from Figure \ref{fig:5} is detected when the four archival signals at 982.002 MHz were detected, just like blc1, and is not detected when the archival signals are not present (see Supplementary Discussion 1.4). 

This interpretation provides an explanation for why the signal was appearing in a part of the spectrum reserved for aviation and navigation: the source was not intending to transmit in that frequency region, and the signal was instead generated by the interaction of electronics within the transmitter, the receiver, or both. In this case, the underlying signals that intermodulate to cause the lookalikes are likely from outside Murriyang's receiver system. Signals are digitized using three analog-to-digital cards inside the telescope focus cabin, with further processing done in RF-tight cabinets in the telescope tower \cite{hobbs2020UWL}. As the lookalikes span across all three analog-to-digital cards, they are unlikely to be spurious signals generated within the receiver's digital systems.

blc1 cannot be the original signal because it is two orders-of-magnitude weaker than the strongest signals in the set and it is not seen in the off-panels, which is not replicated across the set. Evidently, blc1’s duty-cycle or variability tracks the observing cadence on ProxCen, leading to the apparent localization on the sky. If blc1 was always ``on'' at its brightest power, it would have been detected in all off-sources. Supplementary Figures 7 and 9 reveal a range of inter-panel brightness behaviours for the lookalikes: some appeared in every panel, while some were as faint as blc1 and missing from all but the first panel. \texttt{turboSETI}, by the thresholding and on-off selection mechanisms, selected the most interesting signal from a set of potentially hundreds of intermodulation products. These large numbers speak to the particular pathology of this case, as this behaviour had never been seen in over a year of \ac{UWL} observations. 

\ac{RFI} environment analysis is unfortunately complex, and the \ac{RFI} environment around most astronomical radio facilities is not well-characterized at the frequency resolutions used for \ac{SETI} work. In the case of blc1, the situation is more complicated still, with mixing products that obscure the frequency and character of the original, individual interferers. It is possible that we could untangle the origins of this interference. However, as the goal of this study was to determine whether blc1 had an Earth-based or interstellar origin, we find that this is appropriate for future work.

\section{Creation of a Technosignature Verification Framework}

blc1 is the first signal-of-interest from the Breakthrough Listen program that required extensive signal verification to be undertaken. This case study led to a novel signal verification ``toolkit'' for future \ac{SETI} signals-of-interest. Similar frameworks have been applied to searches for fast radio bursts \cite{Foster2018} and gravitational waves (e.g., \cite{abbott2016}), but prior \ac{SETI} programs relied heavily on re-observation alone, without the application of a thorough checklist (e.g., \cite{horowitz, setihome}) We propose the following verification checks for narrowband technosignature signals-of-interest, once known astrophysical origins have been ruled out:

\begin{enumerate}
    \item Verify that all instrumentation was functioning correctly.
    \item Verify that the signal-of-interest was not present in the off-source observations at a lower \ac{S/N} threshold.
    \item Check for catalogued \ac{RFI} at the same frequency at the observatory where the signal-of-interest was discovered.
    \item Compare the drift rate evolution of the signal-of-interest to known accelerational and electronic drifts from human-made technology.
    \item Compare the drift rate evolution of the signal-of-interest to the expected drift rates and periods in the target system and the solar system.
    \item Search for other potential instances of the signal-of-interest in archival data from the same observatory.
    \item Search for similar signals at other frequencies within the observation in which the signal-of-interest was detected. If found, determine if these signals have characteristics consistent with \ac{RFI}.
    \item Qualitatively, or, if possible, quantitatively, assess whether similar signals (from Steps 6 and 7 above) are generated by the same phenomenon as the signal-of-interest.
    \item If other signals \textit{are} from the same phenomenon as the signal-of-interest, determine whether they are \ac{RFI} using the off-source observations, or, if necessary, Steps 1--8.
    \item Re-observe the target with both the same instrument and other instruments to attempt to re-detect the signal-of-interest.
\end{enumerate}

The checklist as written is appropriate for persistent, narrowband technosignature searches with single-beam, single-dish telescopes. The procedure can be applied to multi-beam instruments by using other beams as ``reference pixels'' for the on-source beam, instead of nodding to an off-source position. In addition, re-observation (Step 10) may be performed earlier in the checklist if economical, especially in cases where the signal-of-interest would be expected to be periodic or transient on long timescales (i.e., synchronized with the transit of an exoplanet).

\section{Future Work}

blc1 has underscored that, when practicable, simultaneous observations of potential technosignatures should be conducted at two different observing sites simultaneously. For \ac{ProxCen}, simultaneous observing could be accomplished with MeerKAT (e.g., \cite{jonasMeerKAT2009}) and Murriyang, whose receivers share about 1\,GHz of frequency overlap including the 982\,MHz region of blc1. There is a $\sim$4.5 hour window where the source can be observed simultaneously between the two sites. While re-observations are resource-intensive, they are also scientifically meritorious in their own right. \ac{ProxCen} is still a uniquely fascinating \ac{SETI} target for all of the reasons described in the Introduction, and it will only become better characterized by future, astrobiology-oriented studies. 

We are also investigating other ways to further characterize blc1 in both the hardware and software components of the pipeline. Other Murriyang observers are monitoring for interference near 982\,MHz, which could help us identify position-dependent \ac{RFI}. To characterize aliasing behaviour within the receiver, the sky signal could be excluded by covering the feed with cryogenically-cooled RF-absorbing material which can be treated as a thermal blackbody (known as a ``cold load''). This poses engineering and logistical challenges, but may allow us to perhaps identify the different components whose mixing produces the intermodulation product at 982\,MHz. To understand the particular pathology of variability patterns such as blc1's, we could perform noise-injection testing on the lookalike population. Finally, this case study implies that taking data during slews in future single-dish \ac{SETI} observations could help us better understand signal localization and sidelobe behaviour in future programs.

\section{The impact of blc1}

While we attribute blc1 to \ac{RFI}, the benefits from its unique analysis will inform searches for years to come. This paper outlines a ``checklist'' that provides thorough next-steps for this type of signal. We have developed new software which will be incorporated into \ac{BL}'s analysis packages (e.g., \texttt{blimpy} and \texttt{turboSETI}), such that future signals-of-interest can be assessed more quickly and efficiently.

The detection of blc1 shows the success of the \ac{BL} signal detection pipeline. From 26 hours of observations over billions of channels, the \texttt{turboSETI} algorithm was able to retrieve a set of potential signals-of-interest. blc1 was then easily identified from this set upon visual inspection of the software outputs. 

Conversely, this signal-of-interest also reveals some novel challenges with radio \ac{SETI} validation. It is well-understood within the community that single-dish, on-off cadence observing could lead to spurious signals-of-interest in the case where the cadence matches the duty cycle of some local \ac{RFI}. blc1 provided the first observational example of that behaviour, albeit in a slightly different manner than expected (variation of signal strength over position and time, which changed for each lookalike within the set). This case study prompts further application of observing arrays, multi-site observing, and multi-beam receivers for radio technosignature searches. For future single-dish observing, we have demonstrated the utility of a deep understanding of the local \ac{RFI} environment. To gain this understanding, future projects could perform omnidirectional \ac{RFI} scans at the observing site, record and process the data with high frequency-resolution instrumentation such as the various \ac{BL} backends, and then use narrowband search software such as \texttt{turboSETI} to obtain a population with which to characterize the statistics (in frequency, drift, power, duty-cycle, etc.) of local \ac{RFI}.

Finally, blc1 encourages us to continue working at logistical challenges that have traditionally vexed large, radio \ac{SETI} efforts. In a project with an incredibly high rate of data inflow, how do we work towards data analysis in real-time? When raw voltage data is exceedingly memory-intensive to store, how do we decide pre-analysis which observations may need additional analysis in the raw voltage products? For example, one solution to the data-storage challenge is to store only ``postage stamps'' of events with limited time and frequency ranges. Here we have learned that neglecting to consider the entire operable bandwidth of a receiver can have serious consequences: for example, losing the context that can be used to show a signal-of-interest is \ac{RFI}. 

\section{Data Availability Statement}

All data used in this manuscript are stored as high-resolution filterbank files, which are available through the Breakthrough Listen Open Data Archive at \url{seti.berkeley.edu/opendata}. This includes all observations from the original observing campaign in April 2019, as well as the reobservations in November 2020, January 2021, and April 2021. 

\section{Code Availability Statement}

The software tools used to read these files (I/O) and perform the narrowband search are publicly available at \url{https://github.com/UCBerkeleySETI/blimpy} and \url{https://github.com/UCBerkeleySETI/turbo_seti}.

\section{Acknowledgements}
Breakthrough Listen is managed by the Breakthrough
Initiatives, sponsored by the Breakthrough Prize Foundation. The Murriyang radio telescope is part of the Australia Telescope National Facility which is funded by the Australian Government for operation as a National Facility managed by CSIRO. We thank the staff at Murriyang for their observational support. Shane Smith and Steve Croft were supported by the National Science Foundation under the Berkeley SETI Research Center REU Site Grant No. 1950897. We thank Richard Elkins and Luigi Cruz for help with development and debugging of \texttt{turboSETI}.

\section{Author Contributions}

SZS led the data analysis and wrote the manuscript. SS and DCP uncovered blc1 and analyzed data. DD and BL ran simulations and analyzed data. DC, SC, VG, HI, ML, DM, CN, KP, AVPS, and CW assisted with data analysis, scientific interpretation, and manuscript revision. AZ obtained the data via commensal observing. JD and SPW reviewed the manuscript.

\begin{acronym}
\acro{SETI}{Search for Extraterrestrial Intelligence}
\acro{S/N}{signal-to-noise ratio}
\acro{RFI}{Radio Frequency Interference}
\acro{BL}{Breakthrough Listen}
\acro{ProxCen}{Proxima Centauri}
\acro{UWL}{Ultra-Wideband Low receiver}
\acro{NEO}{Near-Earth Object}
\acro{ETI}{Extraterrestrial Intelligence}
\end{acronym}

\end{document}


\section*{Supplementary Discussion}

\subsection*{Evaluating blc1 on the Rio Scale 2.0}
\label{ssec:rio}

A challenge in discussing potential signals-of-interest in \ac{SETI} is communicating the likelihood and significance of a discovery of \ac{ETI}. The Rio Scale \cite{almar2000discovery} and the revised Rio Scale 2.0 \cite{forgan2019rio} and ranking systems designed to mitigate this challenge. Note that the Rio Scale is built to quantify the novelty and importance of a signal, but it does not provide a framework with which to analyse and determine if a signal is interference. Here we assess blc1 using the Rio Scale 2.0 to provide context for the public for this signal-of-interest.

Using the online Rio Scale interface\footnote{\url{https://dh4gan.github.io/rioscale2/}}, we find that the phenomenon is likely to be instrumental or anthropogenic in origin, largely due to its \ac{RFI}-like form and lack of repeatability, with A (is it amenable to study and repeat observation?), B (are we sure it is not instrumental?), and C (how likely is it to be a technosignature?) values of 6, 8, and 1 respectively. This maps well with conclusions from the main text of this paper. We can then use the A, B, and C values to calculate the warranted journalistic interest $J$ and the likelihood that the signal has an interstellar origin $\delta$ from \cite{forgan2019rio}:

\begin{equation}
    J = A + B + C - 20
\end{equation}

\begin{equation}
\delta = \frac{10(J-10)}{2}
\end{equation}

Here, the J value is 0, despite this being an exciting case study into the challenges of a technosignature search. The Q value of the signal, calculated separately to indicate the depth of investigation that we could pursue \textit{if} there was an ETI transmitting from \ac{ProxCen}, is relatively high, at 7. However, $\delta$, the likelihood that this signal has an interstellar origin, is 0.00001, leading to a final Rio Score score of $R = Q \times \delta = 0$.

Unfortunately, the Rio Scale does not actually give us much information about signals like blc1. There is a lack of granularity, noted in the original paper, for events that score between 0 and 1 in Rio Score. We propose that a new scaling system, perhaps logarithmic, should be developed to provide appropriate context and comparison for signals at blc1's level of interest.

\subsection*{Proxima Centauri is not alone in the beam}

In single-dish radio astronomy, the size of the beam on the sky is much larger than the target's angular extent, except for extended/diffuse sources. The low frequencies used in these data translate to approximately a 15\arcmin\ beam, which, according to GAIA DR2 \cite{gaia2018vizier} contains $\sim 22000$ known objects along this sight-line. This number is in fact a lower limit, as it does not account for extragalactic objects in the beam. This underscores the fact that, if we were to find a signal coming from interstellar distances, we would have to perform follow-up observations (assuming the source was continuous or had a relatively high duty-cycle) to determine where in the beam the source was coming from. In this example, while the beam was centered on \ac{ProxCen}, it is more accurate to say that the detection was made ``in the direction of \ac{ProxCen}''.

\subsection*{Expectations for the Proxima Centauri System}
  
Our analysis is mostly agnostic to the parameters, number, or habitability of planets within the triple-star system of Alpha Centauri; however incorporating this information provides a good case study for the benefits and limitations of including exoplanetary parameters in a drift analysis. Proxima\,b is a 1.3\,M$_{\oplus}$ exoplanet with an 11.2\,d period, and was detected using the ESO HARPS spectrograph \cite{anglada-escude2016terrestrial, mascareno_2020}. It has received the majority of astrobiological attention in the system, as it resides at 0.0485\,AU from its host star, and has an equilibrium temperature of 234\,K \cite{anglada-escude2016terrestrial}. There are additional planets in the \ac{ProxCen} system. In 2019, a second planet, Proxima c, with a minimum of 7 M$_{\oplus}$ and a 5.21 year period was confirmed via radial velocities from HARPS spectra data \cite{Damasso_2020}. A third planet candidate signal, which would be Proxima d, was also detected in \cite{mascareno_2020} with a 5.15 day period and a minimum mass of 0.29 M$_{\oplus}$. Given \ac{ProxCen}'s habitable zone range of $\sim$ 0.03--0.09 AU \cite{Alvarado_2020, Kane_2012}, the Proxima c candidate would be far outside of the habitable zone \cite{Damasso_2020}, while the Proxima d candidate would likely experience extreme weather conditions due to its proximity to \ac{ProxCen} \cite{Alvarado_2020}. Proxima d has yet to be confirmed, and no moons or rings have been confirmed around any planet in the system so far. Still, after finding a bright signal from infrared SPHERE images, \cite{Gratton_2020} speculated the existence of rings or dust around Proxima c. 

Given our uncertainty about the number and characteristics of planets in the system, it is difficult to constrain the drift rates that we would expect from a transmitter in the system. Nevertheless, we can apply an order-of-magnitude approach for the currently-known planets. The orbital motion of Proxima b, due to its short orbital period, can produce drift rates up to $10^2$ times higher than that observed in blc1 when it is closest to the Earth. Proxima b's rotational speed is unknown, but if we set the lower and upper limits to the theoretical values when assuming that the planet is tidally-locked (lower) and assuming that the planet is rotating at its break-up speed (upper)\footnote{Using the methodology described in \cite{sheikh2019choosing}}, then the rotation of Proxima b could produce values from $10^{-1}$--$10^3$ of the observed blc1 drift rate. The consensus in the literature is that Proxima b should be tidally-locked (e.g., \cite{ribas2016prox}), and thus it is likely that rotation alone could not cause a large contribution to a drift rate from a transmitter on its surface --- the opposite of the situation of drift rate contributions of the Earth. Proxima c is likely a gas giant, and thus cannot host a transmitter on its surface (no component from rotational motion). In addition, its orbital motion would be an order-of-magnitude too low to produce drifts on the order of blc1.

Drift rate analyses like these can be useful as a heuristic tool to determine consistency with expected drift magnitude and sign, but in practice, they cannot be used to rule out a potential signal-of-interest. Even if we assume that any drifts that remain after a barycentric correction must be accelerational, it is impossible to determine \textit{a priori} exactly which drift rates we would expect at which times. The drift rate vs. time of a signal is dependent on which planet the signal is coming from, whether the transmitter is orbital or on a planetary surface, and the transmitter's latitude/angle of orbit relative to Earth. This doesn't even account for transmitters that are orbiting \ac{ProxCen} itself, orbiting a Lagrange point in the system (such as SOHO in the Solar System, e.g., \cite{domingo1995soho}) or placed on moons. In addition, we cannot neglect drifts from electronic oscillators, either unintentional or intentional. A signal from an \ac{ETI} may be electronically de-drifted for the rotational and/or orbital motion of the host system, de-drifted for the receiver's system (not necessarily Earth), or intentionally sweeping in frequency for some unknown functional reason.

\subsection*{Other appearances of blc1 also show intermodulation lookalikes}

We used the analysis above to provide another line of evidence that the archival 982.002 MHz signals (some of which are clearly RFI) are definitively related to blc1. We searched for lookalikes with the same spacings as in Figure 5. We find lookalikes at 712\,MHz and 856\,MHz in 4 of the 5 archival signals, while the 1062\,MHz lookalike is only present in 1 of them. This is not surprising, given that 1062\,MHz is the faintest lookalike in the blc1 set; the frequency comb interference is stronger at these higher frequencies; and that the entire cadence is washed out by broadband interference for one of the archival signals. To double-check this interpretation, we searched for signals at these frequencies from about a month of \ac{UWL} observations surrounding the \ac{ProxCen} campaign where we do \textit{not} see blc1 and find that, for the majority of observations, we do not see any activity at any of these frequencies either (the exception being three faint appearances of 712\,MHz over the entire month). 

\section*{Supplementary Methods}

\subsection*{Determining blc1's Fundamental Parameters}

\label{ssec:blc1params}

While \texttt{turboSETI} provides a first-order best-fit start frequency and drift rate, we needed to determine blc1's fundamental parameters to higher precision in order to perform the more complex analyses detailed in the main text. We determined that the original signal on 2019 April 29 (blc1) appeared at \ac{SNR} $> 10$ in five observations. The properties of the five observations are shown in Supplementary Table S1. 

To build Supplementary Table S1, we replicated \texttt{turboSETI}'s search strategy with a de-drifting algorithm. We tested a range of 500 drift rates between 0 and 0.05 Hz/s, informed by the initial \texttt{turboSETI} estimates. At each drift rate, the time rows in the time-frequency-power array were ``rolled'' to de-drift to the test drift rate; an illustration of this method is shown in Supplementary Figure S12. Once the array had been de-drifted, we then summed the spectra in the array over the entire 30\,min observation and searched for a power maximum in frequency space. We assigned the best-fit drift rate to the drift rate that produced the largest power maximum. 

It is unfortunately impossible to assign a statistically-valid error value to these measurements without knowing the underlying distribution. In an idealized case, the distribution of power values around the best-fit drift rate will be symmetric and have more data in the tails of the distribution (excess kurtosis) than a Gaussian. However, because the data are discrete, this distribution will be affected by the interplay of frequency resolution, integration time, and drift rate, and these parameters will cause additional, finer structure within the distribution. This is before the noisiness and presence of \ac{RFI} within the dataset are even considered; the problem of drift rate best-fit errors within quantized data is complex, and has not been treated in the literature, to our knowledge. For the purposes of this paper, we determined a proxy for the drift rate error by fitting a Gaussian to the distribution of maxima in drift rate space. Although a Gaussian does not provide a perfect fit for the reasons described above, we are still able to obtain a relative measure of certainty in the drift rate, based on fundamental, interesting parameters such as the signal's linearity, duration and strength. 

We determined the start time of the signal by dividing each 30\,min observation into 5\,min slices, and choosing the start time of the first slice where the de-drifted signal becomes significant (\ac{SNR} $\gtrsim 5$) in that observation. We set the start frequencies to the location of the maximum in frequency space for the best-fit drift rate. Finally, we determined the \ac{SNR} by dividing the summed power in the signal region over the entire 30 minute observation by the root-mean-square of power in the non-signal region for the de-drifted (best-fit drift rate) array. 

We can use the average \ac{SNR} from Supplementary Table S1 to calculate blc1's apparent brightness. The \ac{UWL} system equivalent flux density (SEFD) of blank sky at 982 MHz is $\sim$38\,Jy. Thus, with an average \ac{SNR} of $\sim 18$, we can get an order-of-magnitude source brightness of 
\begin{equation}
    S_{\nu} \sim \frac{\textrm{SNR} \times \textrm{SEFD}}{\sqrt{t_{int} \times \Delta \nu}}
    = \frac{18 \times 38 \textrm{Jy}}{\sqrt{1800 \textrm{s} \times 4 \textrm{Hz}}}
    \sim 8 \textrm{Jy}
\end{equation}

\subsection*{Drift Rates from Reflections of Solar System Objects}

Although blc1 was detected during an observation of \ac{ProxCen}, we wanted to consider whether it could instead be related to a Solar System object that happened to be in or near the beam. This hypothetical object could be an extraterrestrial technosignature itself, e.g. a transmitter \cite{Bracewell1960}, beneficial from the perspective of a hypothetical \ac{ETI} because it would require orders-of-magnitude less power than an interstellar transmitter (e.g., \cite{Freitas1985,Benford2019}). On the other hand, a Solar System object may have caused blc1 via an \ac{NEO} acting as a reflector of an Earthbound radio transmitter.

The main difficulty with either explanation is that \ac{ProxCen} is far off the ecliptic plane ($b = -44\fdg8$), which is inconsistent with an object on (or in orbit around) a major planet. An interplanetary transmitter could easily be bright enough to be detected far into the sidelobes, of course, but this would not explain the null detections when observing the calibrators; without this constraint, a simpler explanation would surely be an anthropogenic satellite. Nonetheless, a Solar System object could appear in the vicinity of \ac{ProxCen} if it either has a large inclination or is a \ac{NEO} making a close approach to Earth.

We checked whether any known minor body was near \ac{ProxCen} using the Jet Propulsion Laboratory's online ``What's observable?'' tool (\url{https://ssd.jpl.nasa.gov/sbwobs.cgi}).  We searched for objects in the \ac{RA} range \hms{13}{44}{43} to \hms{15}{14}{43} and \ac{Dec} range \dms{-57}{40}{46} to \dms{-67}{40}{46}, which includes everything within $5^{\circ}$ of \ac{ProxCen}, at UT 2019 April 29 15:30 as observed from Murriyang.  53 objects were returned, but most were several degrees away from \ac{ProxCen} at the time.  For comparison, Murriyang's beam width at 982\,MHz is $\sim \lambda / (64{\rm\,m}) \approx 0\fdg27$.

One object stood out, however: 2019\,GG$_3$, which was within $1^{\circ}$ of \ac{ProxCen} at the time (Supplementary Figure S13). During the blc1 observations, 2019\,GG$_3$ was outside the main lobe of Murriyang, although it would pass within its footprint a couple of hours later. Coincidently, further investigation revealed that 2019\,GG$_3$ had passed within 0.07 AU of Earth just four days prior, $< 1$\% of its Minimum Orbit Intersection Distance.  JPL's Small-Body Database lists that the only close encounter between 2019\,GG$_3$ and Earth was that April 25 encounter (\url{https://ssd.jpl.nasa.gov/sbdb.cgi?sstr=2019\%20GG3;orb=0;cov=0;log=0;cad=1\#cad}). 

We evaluated whether the drift rate was compatible with a transmitter on 2019\,GG$_3$ using its radial velocity as observed from Murriyang, as calculated using JPL's HORIZONS ephemerides system (\url{https://ssd.jpl.nasa.gov/horizons.cgi}). The predicted drift rate (Supplementary Figure S14) is clearly incompatible with the observed drift rate (with $\chi^2 \sim 1100$) -- although the qualitative shape is correct, it has the wrong sign and is about half the magnitude expected.  Thus, a constant frequency transmitter on 2019\,GG$_3$ itself was not responsible for blc1.  We did not consider rotation on 2019\,GG$_3$ as a possible contributor to the drift rate. 

Hypothetically, 2019\,GG$_3$ may have reflected an Earthbound transmitter.  The expected returned flux is very small:
\begin{equation}
\label{eqn:Fnu_NEO}
F_{\nu} = 56\ \mathrm{mJy}\ \left(\frac{\mathrm{EIRP}}{1\ \mathrm{TW}}\right)\left(\frac{B}{3\ \mathrm{Hz}}\right)^{-1} \left(\frac{\sqrt{D_R D_T}}{0.069\ \mathrm{AU}}\right)^{-4} \left(\frac{A_r}{3000\ \mathrm{m^2}}\right),
\end{equation}
where $B$ is the transmission bandwidth, $D_R$ and $D_T$ are the reflector's distance from Earth, and $A_r$ is the effective cross section for reflection as viewed from Earth.  With an absolute magnitude of $H = 24.2$, 2019\,GG$_3$ has a radius of $9.7/\sqrt{p}$ meters for an albedo $p$.  Thus for an $A_r$ comparable to its geometric area, even the brightest radars on Earth would be incapable of generating a detectable return.  The maximum possible $A_r$ for an object with geometric area $A_g$ can be estimated with Rayleigh's criterion to be $\sim 4 \pi A_g^2 / \lambda^2$, or $\sim 10^3\,\mathrm{km}^2 (A_g / 3000\,\mathrm{m}^2)^2$ at 982\,MHz.  This would allow a return of 0.28 Jy (with a transmitter EIRP of 1 TW, a channel bandwidth of 3 Hz, a distance of 0.069 AU, and an effective radar cross-section of order $\pi/4$ (60 m)$^2$). It would require a rather contrived reflector, essentially a flat sheet oriented precisely to bounce the transmitter's signal into Murriyang.

A further argument against 2019\,GG$_3$ being a reflector comes from the drift rate.  The predicted drift rate of a reflection is the drift rate due to the sums of the radial acceleration between transmitter and reflector and that between the reflector and receiver.  Given position vectors $\mathbf{r_T}$ for the transmitter, $\mathbf{r_S}$ for the apparent source (reflector), and $\mathbf{r_R}$ for the receiver:
\begin{equation}
\dot{\nu} = \frac{\nu}{c} \left[ \frac{(\mathbf{\ddot{r}_S} - \mathbf{\ddot{r}_T}) \cdot (\mathbf{r_S} - \mathbf{r_T}) + v_{TS}^2 - v_{TS_{\rm rad}}^2}{|\mathbf{r_S} - \mathbf{r_T}|} + \frac{(\mathbf{\ddot{r}_S} - \mathbf{\ddot{r}_R}) \cdot (\mathbf{r_S} - \mathbf{r_R}) + v_{RS}^2 - v_{RS_{\rm rad}}^2}{|\mathbf{r_S} - \mathbf{r_R}|}\right] ,
\end{equation}
where $v_{xy} = |\mathbf{\dot{r}_x} - \mathbf{\dot{r}_y}|$ is the relative speed of $x$ and $y$ and $v_{xy_{\rm rad}} = (\mathbf{\dot{r}_x} - \mathbf{\dot{r}_y}) \cdot (\mathbf{r_x} - \mathbf{r_y}) / |\mathbf{r_x} - \mathbf{r_y}|$ is the radial velocity between $x$ and $y$.  When both the transmitter and receiver are Earthbound and the reflector is at a great distance, one can ignore parallax and the drift rate is approximated as:
\begin{equation}
\dot{\nu} \approx \frac{\nu}{c} \left[2 \dot{v}_{S\oplus_{\rm rad}} + \frac{4 \pi^2 R_{\oplus}}{P_{\oplus}^2} \cos \delta_S (\cos \Lambda_R \cos(\alpha_S - \tau_R) + \cos \Lambda_T \cos(\alpha_S - \tau_T))\right] .
\end{equation}
In this equation, $\dot{v}_{S\oplus_{\rm rad}}$ is the radial acceleration between the reflector and Earth's geocenter, $\alpha_S$ and $\delta_S$ are the \ac{RA} and \ac{Dec} of the reflector as viewed from Earth, $R_{\oplus}$ and $P_{\oplus}$ are the radius and sidereal rotation period of the Earth, $\Lambda_T$ and $\Lambda_R$ are the latitudes of the transmitter and receiver, and $\tau_T$ and $\tau_R$ are the sidereal time at the transmitter and receiver expressed in radians.

We modeled transmitters located on a $1^{\circ} \times 1^{\circ}$ equirectangular grid on Earth's surface.  The fit between the measured drift rates and the predicted drift rates are evaluated using the $\chi^2$ value.  No transmitter location does an adequate job explaining blc1's drift as a reflection off 2019\,GG$_3$ (Supplementary Figure S15).  The problem is that the transmitter-asteroid drift must overcome both the negative drift from Murriyang and the negative drift from 2019\,GG$_3$'s recession from Earth.  When considering transmitters for which 2019\,GG$_3$ was above the horizon for blc1's duration, the best fit is $\chi^2 = 538$.

2019\,GG$_3$'s encounter with Earth and its near-transit of \ac{ProxCen} during the blc1 time frame is thus almost certainly just an interesting coincidence.  These considerations do not rule out that an unknown Solar System object was responsible for blc1, although it would need to be small to have gone undetected.

\begin{table}[ht]
    \centering
    \begin{tabular}{c|c|c|c|c|c}
        \hline
        \# & Start Time (UTC) & Start Freq. (MHz) & DR (Hz/s) & DR Error Proxy (\%) & SNR \\
        \hline
        1 & 13:17:35 & 982.00241 & 0.0312 & 15.1 & 12.64 \\
        2 & 13:47:46 & 982.00247 & 0.0303 & 22.8 & 28.09 \\
        3 & 14:43:40 & 982.00257 & 0.0305 & 14.8 & 21.17 \\
        4 & 15:21:41 & 982.00264 & 0.0273 & 22.3 & 17.70 \\
        5 & 17:54:26 & 982.00281 & 0.0112 & 67.0 & 10.18 \\
        \hline
    \end{tabular}
    \caption{Start time and frequency, drift rate DR (with error proxy), and signal-to-noise ratio SNR for the five observations of \ac{ProxCen} which comprise blc1. All five observations have strong (\ac{SNR} $>$ 10) detections, all detections appear to be part of the same phenomenon, and the signal is strongly drifting, with a changing drift rate over time, perhaps indicative of changing radial acceleration or electronically-caused frequency shifting.}
    \label{tab:blc1_properties}
\end{table}

\begin{table}[ht]
    \centering
    \begin{tabular}{|c|c|c|c|}
         \hline
         Fundamental Frequency (MHz) & Harmonic Frequencies (MHz) & $n$th Harmonic \\
         \hline
         108.225 & 865.799927, 1298.699886 & 7, 11 \\ 
         133.200 & 799.199932, 1065.599907, 1198.799896 & 5, 7, 8 \\ 
         144.300 & 865.799927, 1298.699886 & 5, 8 \\ 
         149.850 & 899.099922, 1198.799896 & 5, 7 \\ 
         199.800 & 799.199932, 1198.799896 & 3, 5 \\ 
         216.450 & 865.799927, 1298.699886 & 3, 5 \\ 
         266.400 & 799.199932, 1065.599907 & 2, 3 \\ 
         299.700 & 899.099922, 1198.799896 & 2, 3 \\ 
         399.6 & 799.199932, 1198.799896 & 1, 2 \\ 
         432.900 & 865.799927, 1298.699886 & 1, 2 \\ 
         \hline
    \end{tabular}
    \caption{A set of harmonic sequences that were discovered by the harmonic analysis in Results, defined by their fundamental frequency, the frequencies of the lookalikes that are members of the sequence, and the respective harmonic number $n$. blc1 is not a member of any of these sequences, however, these sequences draw from the same sub-set of lookalikes (with harmonic frequencies at ``x.y99'' MHz), which indicates that at least some of the lookalikes are related in origin.}
    \label{tab:harmonics}
\end{table}

\begin{figure}[ht]
    \centering
    \includegraphics[width=11.5cm]{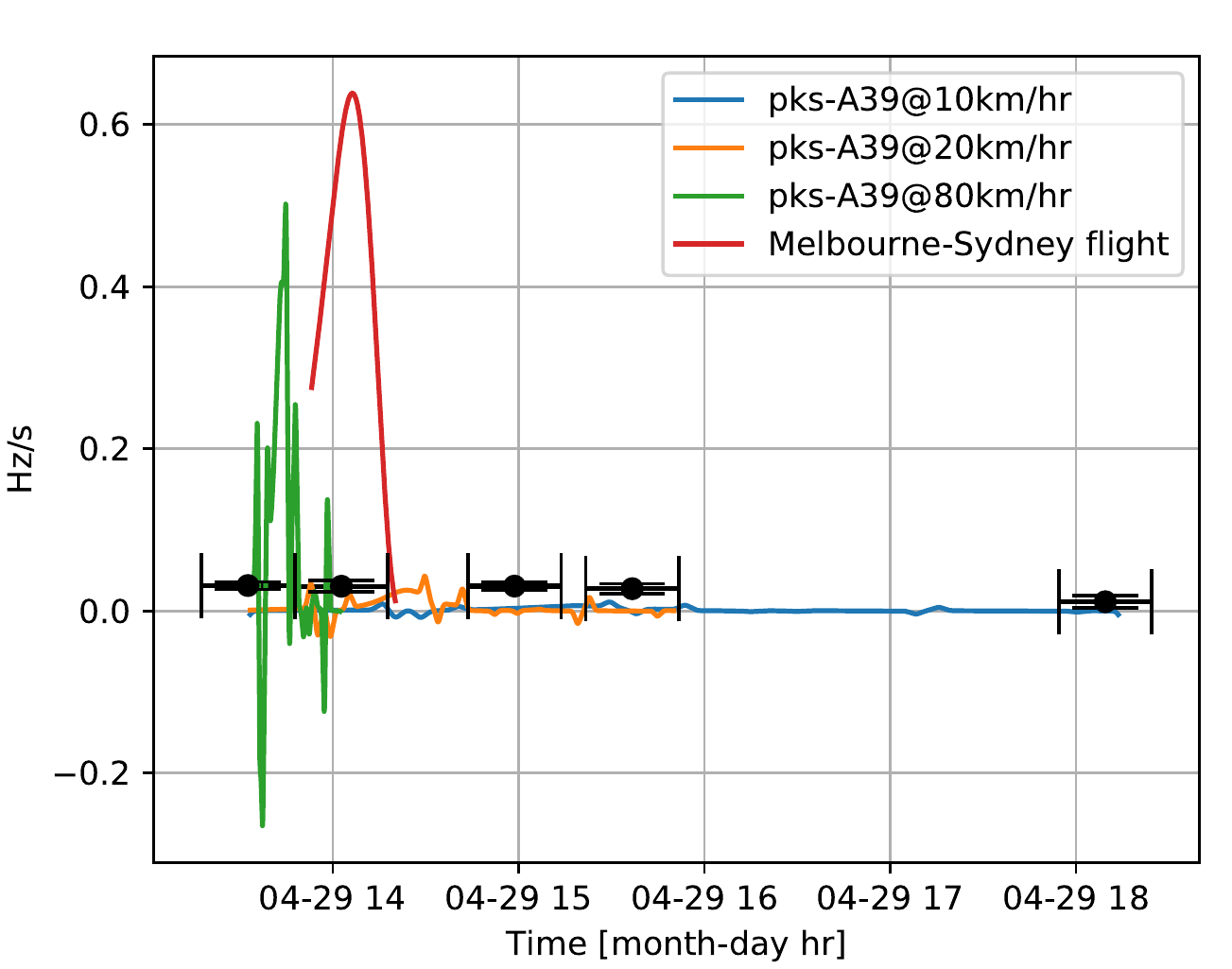}
    \caption{Doppler drift characteristics of a fixed-frequency transmitter on a vehicle travelling along a local highway (A39) at 80 km/hr (green), 20 km/hr (orange), and 10 km/hr (blue), as well as on a plane executing a 850 km/hr great circle flight from Melbourne to Sydney (red, only shown while it is above the horizon).  The black data points show the blc1 data, where the x-axis bars are the time extent and the y-axis are error bars. All ``trips'' start at the first data point, and none can explain the blc1 drift characteristics.
\label{fig:EarthBoundDrift}}
\end{figure}

\begin{figure}[ht]
    \centering
    \includegraphics[width=\textwidth]{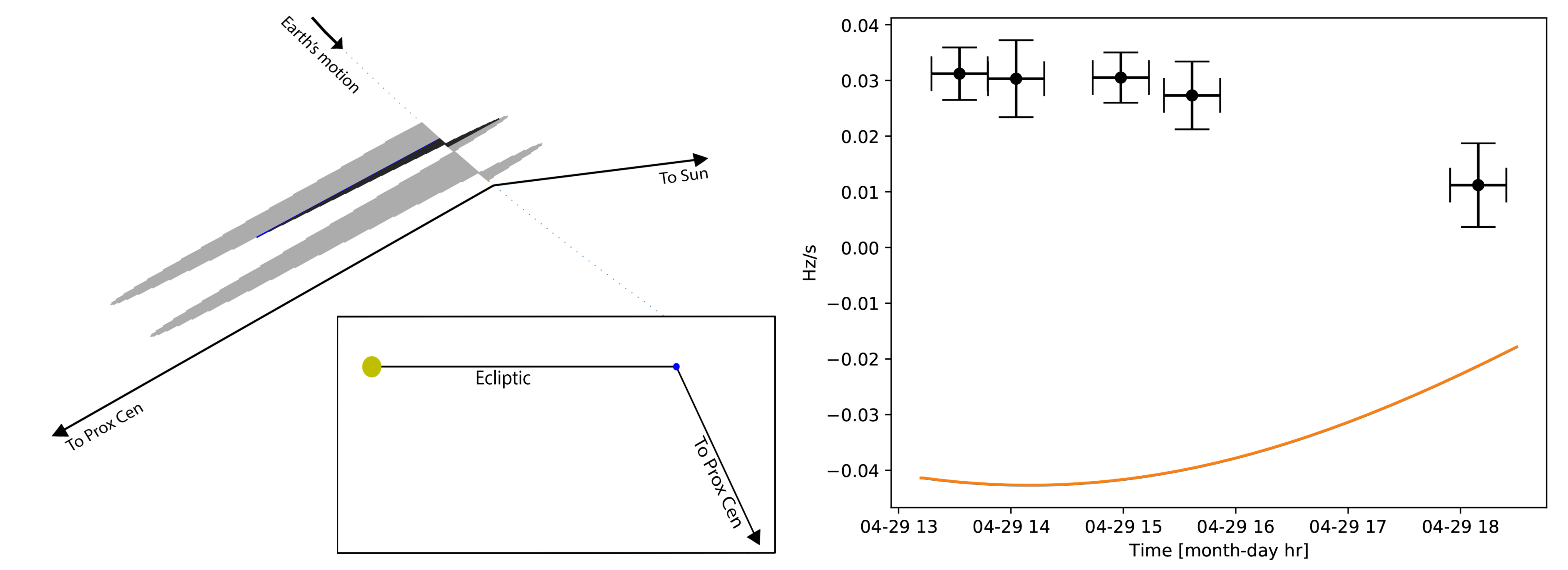}
    \caption{The cartoon on the left shows the geometry of the solar system in the plane of the ecliptic and perpendicular to that plane. The sinusoidal pattern shows the net velocity in the pointing direction, as seen from Parkes, over two days. The black portion is during the appearance of blc1. The orange line in the drift rate vs. time plot on the right shows the result of that motion as residual barycentric/geocentric drift (on this scale both corrections are identical), compared to the actual observed datapoints of blc1. The motion from Earth's rotation produced a drift rate in the direction of ProxCen, over the course of the observation, that is not consistent with the drift we see from blc1.
    \label{fig:BaryGeoDrift}}
\end{figure}

\begin{figure}[ht]
    \centering
    \includegraphics[height=0.92\textheight]{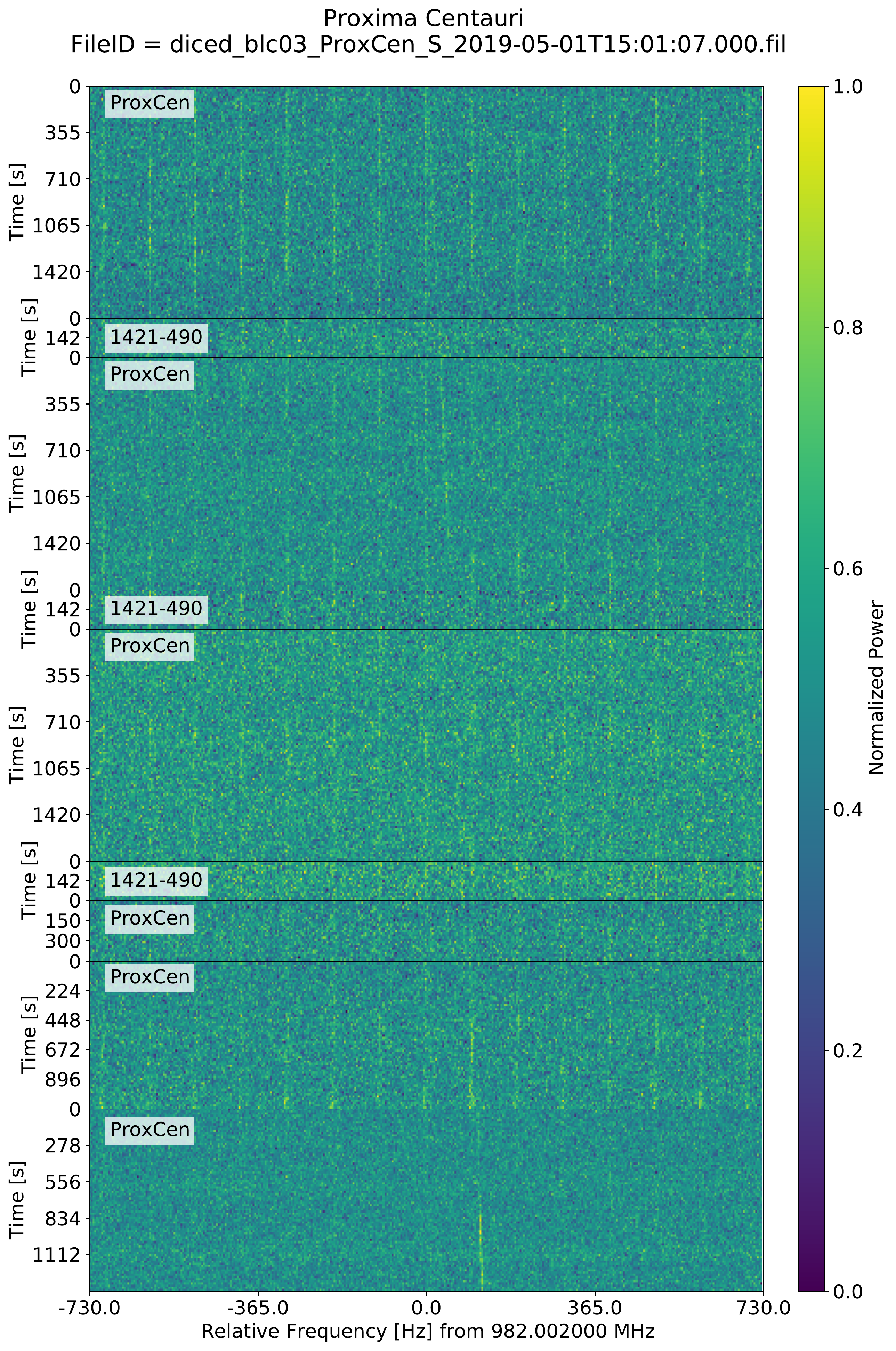}
    \caption{A waterfall plot around 982\,MHz from 2019 May 1, showing a possible blc1 redetection. The \ac{RFI} comb is clearly visible in this observation.}
    \label{fig:blc1c_cadence}
\end{figure}

\begin{figure}[ht]
    \centering
    \includegraphics[height=0.92\textheight]{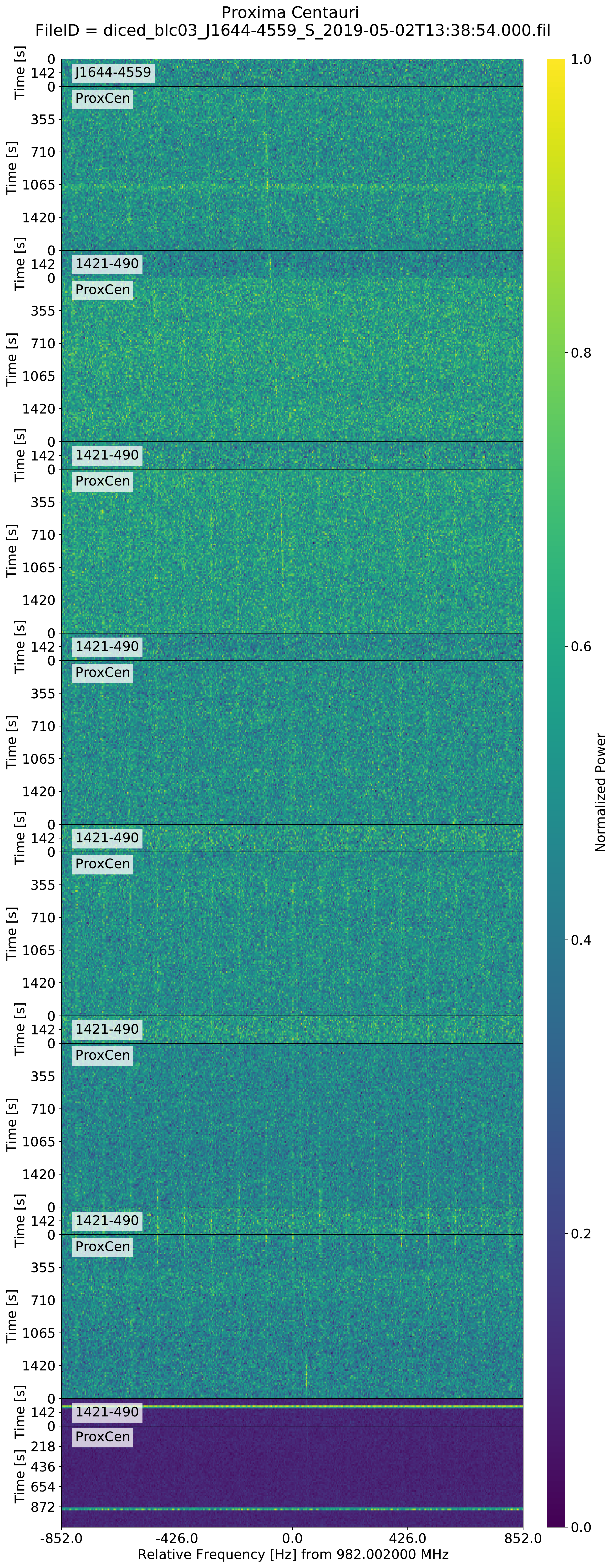}
    \caption{A waterfall plot around 982\,MHz on 2019 May 2, showing a possible blc1 redetection. As with Figure S4, the \ac{RFI} comb is clearly visible in the background as well as broadband \ac{RFI} during the final two observations in the sequence.}
    \label{fig:blc1d_cadence}
\end{figure}

\begin{figure}[ht]
    \centering
    \includegraphics[height=0.92\textheight]{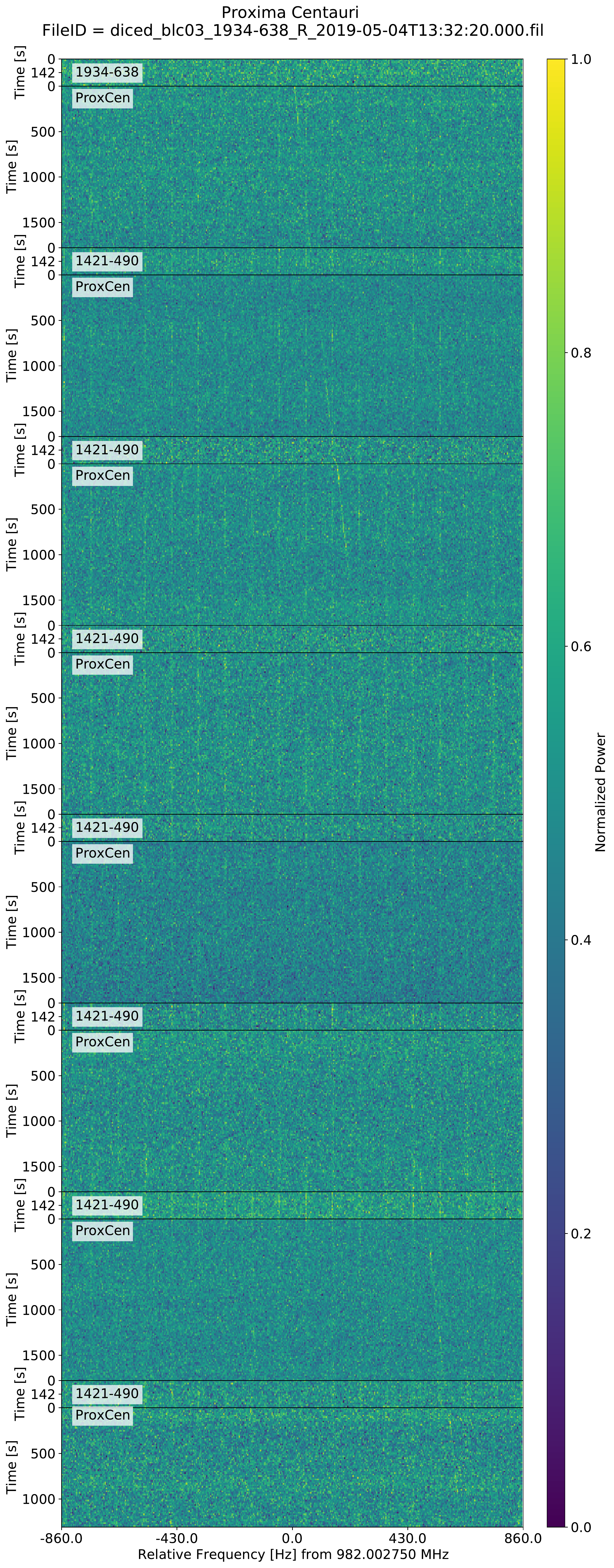}
    \caption{A waterfall plot around 982\,MHz on 2019 May 4, showing a possible blc1 redetection. The possible redetection has a low \ac{SNR} of $\sim$6, comparable with the \ac{RFI} frequency comb.}
    \label{fig:blc1e_cadence}
\end{figure}

\begin{figure}[ht]
    \centering
    \includegraphics[height=0.9\textheight]{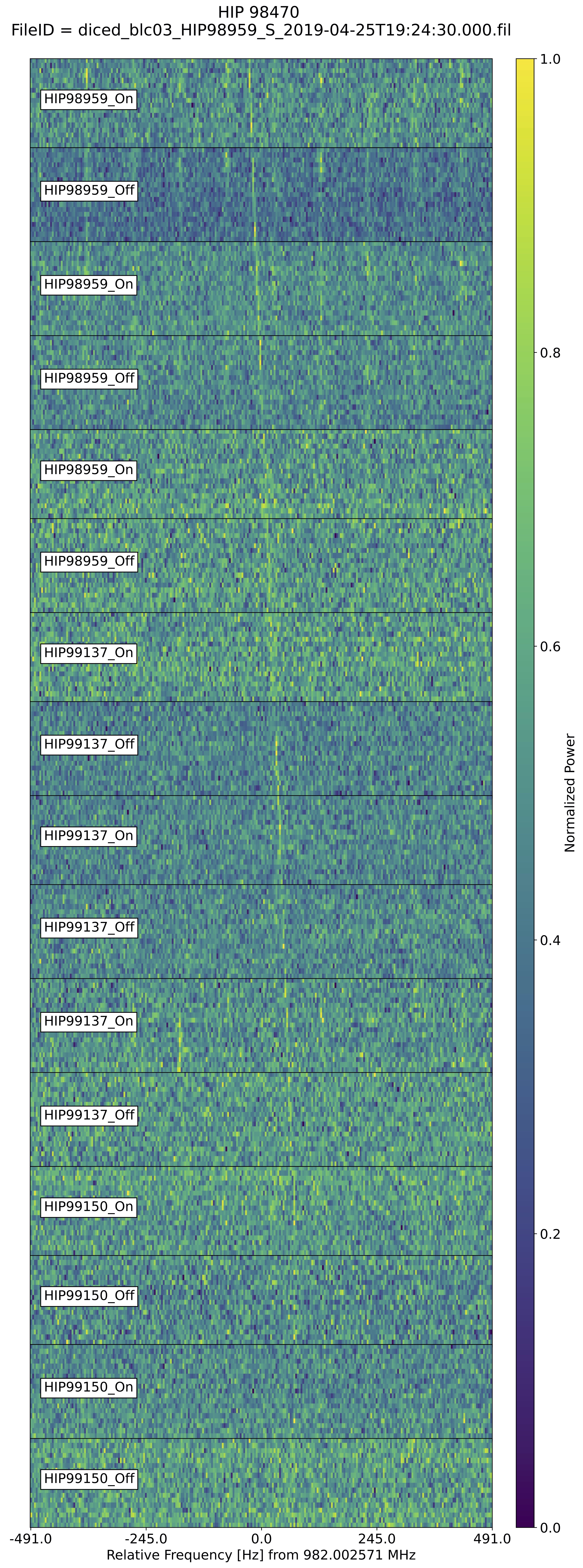}
    \caption{A waterfall plot around 982\,MHz depicting an \ac{RFI} signal detected 4 days prior to blc1, in the direction of HIP 98470 (not part of the \ac{ProxCen} campaign). Note that each panel shows a 5-minute integration, and time still progresses from top to bottom. This signal is also simultaneous with the frequency comb, and appears to have frequency, drift rate, and \ac{SNR} similar to blc1.}
    \label{fig:rfi}
\end{figure}

\begin{figure}[ht]
    \centering
    \includegraphics[height=0.83\textheight]{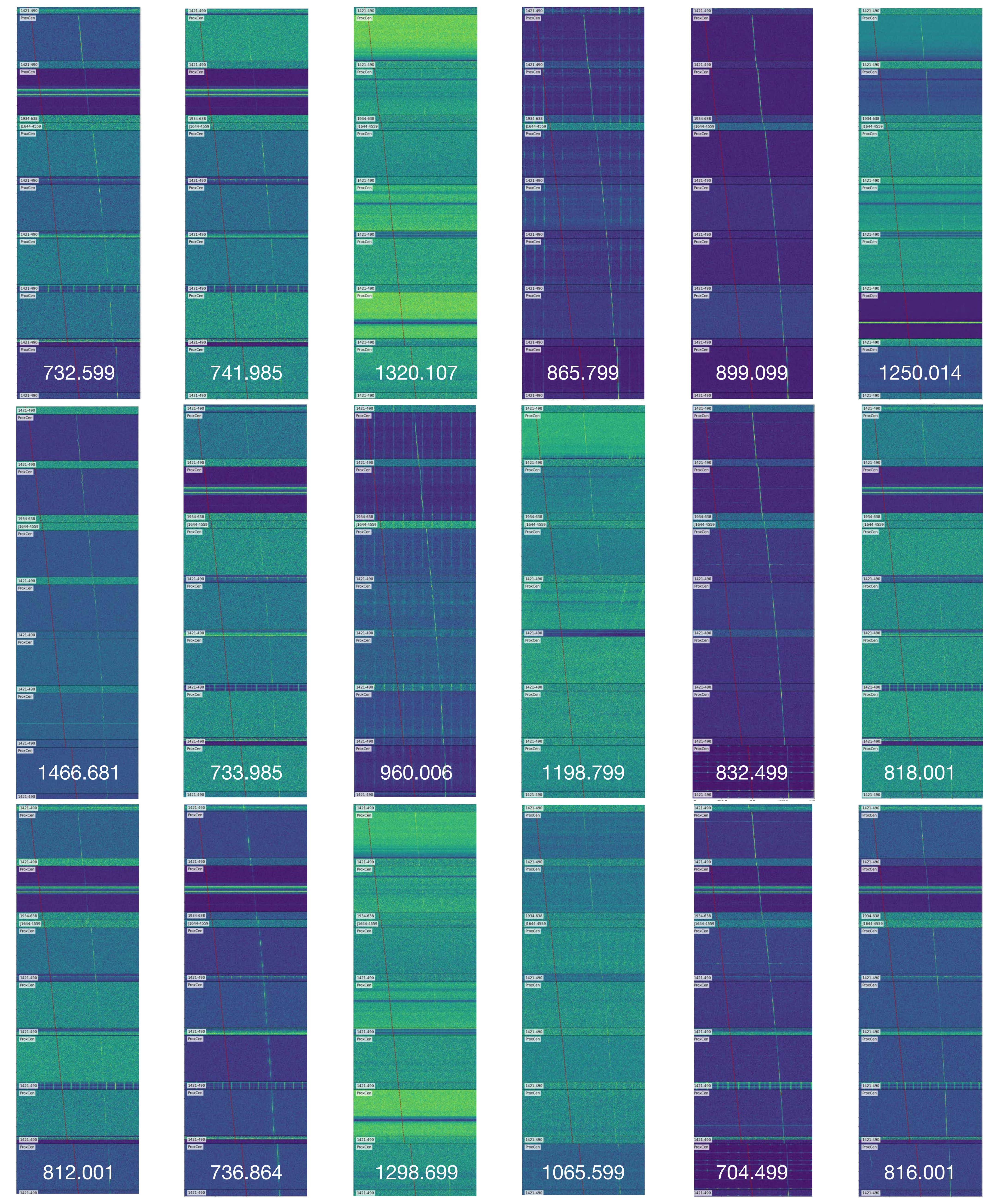}
    \caption{Waterfall plots for 18 of the 36 \ac{RFI} signals, seen in the same observation as blc1, that were found to have the same drift morphology as blc1. All plots cover the same observing time range on the same targets, but at different frequencies (displayed on the bottom of each plot in MHz). These plots are meant to serve as a visual illustration of the sample and thus have had their axes removed; for a quantitative survey of the parameters, see Figure 5. Note the qualitative differences in signal strength, brightness over time, stability over short timescales, and upon close examination, relative ending frequency. These features, along with a lack of a mathematical relation between every member of this set, indicate that these signals probably do not come from a single source.} 
    \label{fig:same_morph}
\end{figure}

\begin{figure}[ht]
    \centering
    \includegraphics[height=0.85\textheight]{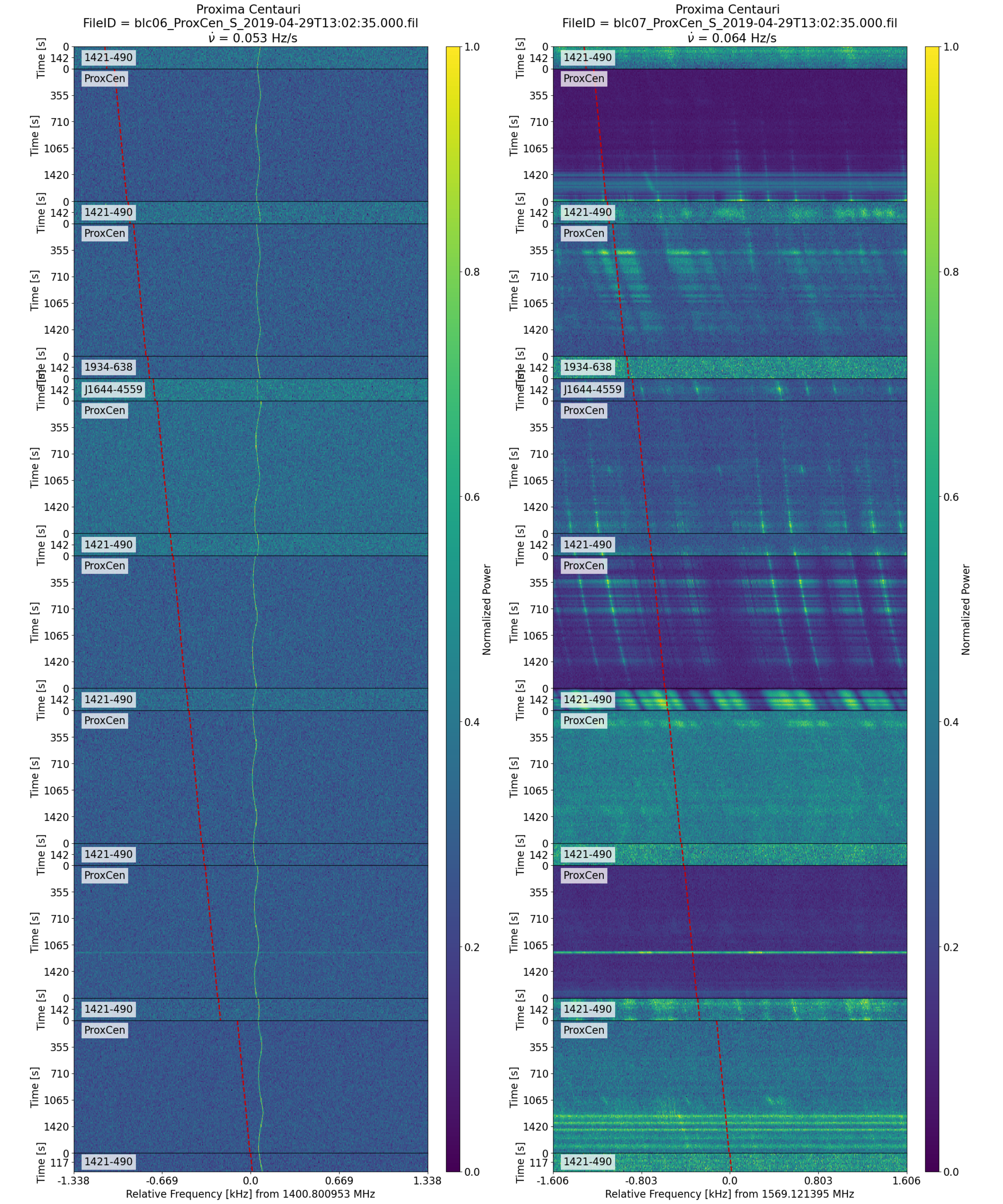}
    \caption{Two morphologically distinct examples from the set of 76 signals with similar drift rates to blc1 in the first panel that \textit{did not} have the same drift morphology as blc1. These do not appear to be produced by the same mechanism as the lookalikes in Figure \ref{fig:same_morph}.}
    \label{fig:non_lookalikes}
\end{figure}

\begin{figure}[ht]
    \centering
    \includegraphics[height=0.85\textheight]{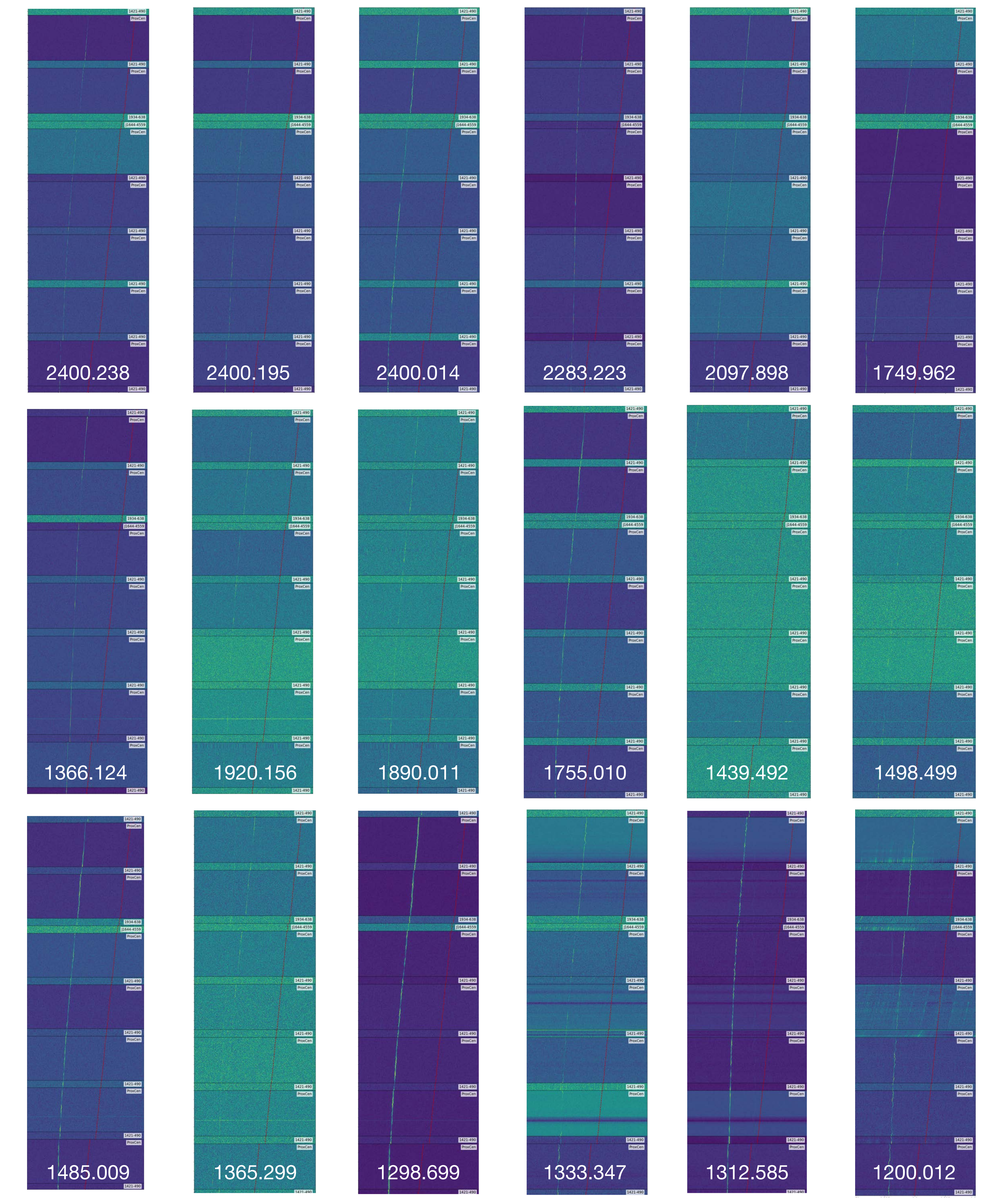}
    \caption{Waterfall plots for 18 of the 27 \ac{RFI} signals, simultaneous to blc1, that were found to have the same drift morphology as blc1 but with negative drift rates. All plots cover the same observing time range on the same targets, but at different frequencies (displayed on the bottom of each plot in MHz). These plots are meant to serve as a visual illustration of the sample and thus have had their axes removed; for a quantitative survey of the parameters, see Figure 5. Again we see significant differences in various signal parameters.} 
    \label{fig:mirror_morph}
\end{figure}

\begin{figure}[ht]
    \centering
    \includegraphics[width=\textwidth]{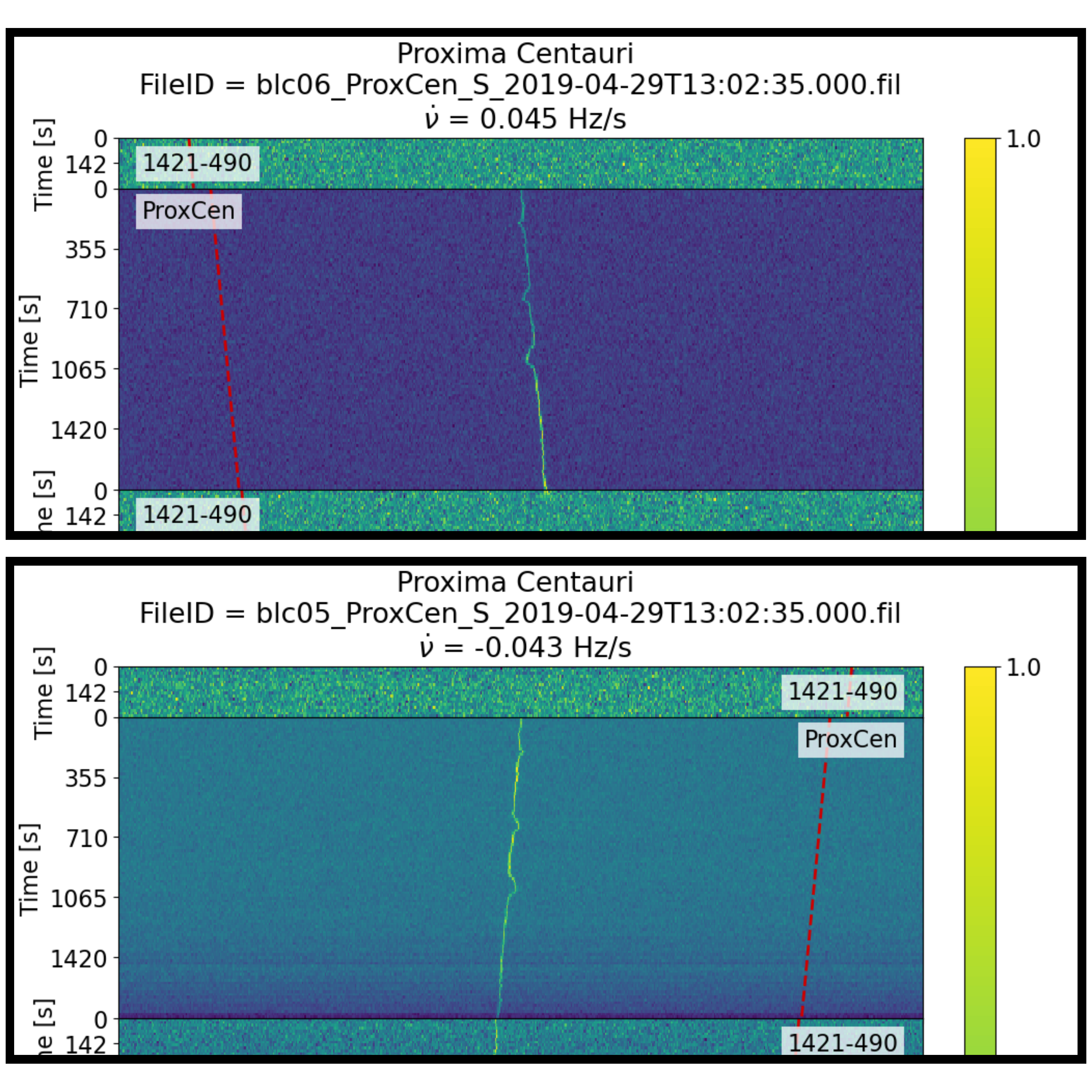}
    \caption{The first panel of a \ac{ProxCen} observation of a blc1 lookalike (top) and blc1 mirrored lookalike (bottom) that show the exact same frequency vs. time morphology, but inverted along the frequency axis. The ``triple feature'' (TF) name was given based on the three wiggles, to the left in the top panel and to the right in the bottom panel, which created a distinctive morphology only displayed by a small sub-set of lookalikes.}
    \label{fig:TF}
\end{figure}

\begin{figure}[ht]
    \centering
    \includegraphics[width=0.97\textwidth]{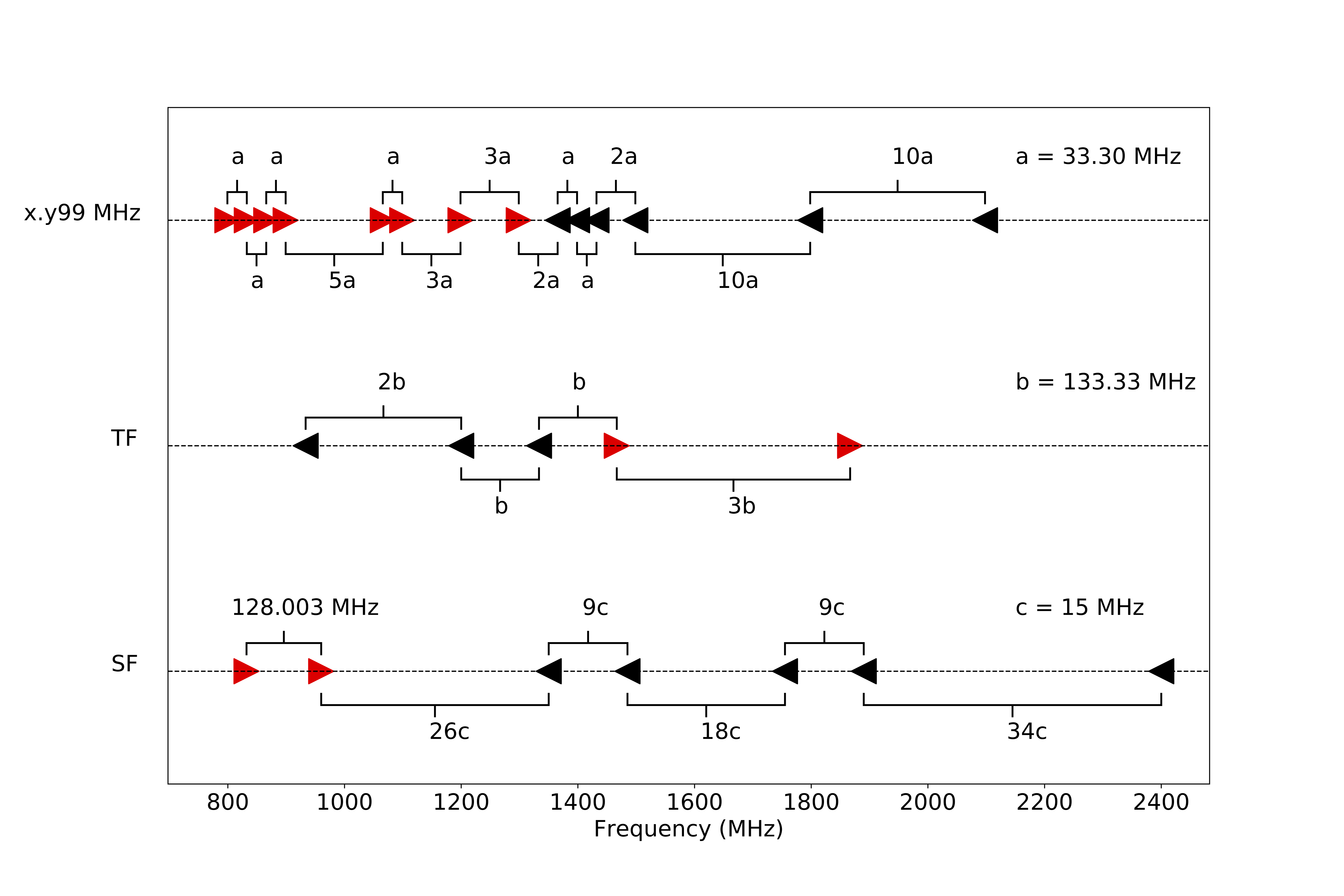}
    \caption{Three sets of lookalikes, including both mirrored and non-mirrored members, with spacings that are integer multiples of common clock oscillator frequencies. Red (right-facing arrow) indicates a positive lookalike while black (left-facing arrow) indicates a negative lookalike. The top sequence, labelled ``x.y99 MHz'', is a set of 14 signals separated by integer multiples of 33.3\,MHz, found via similarities in their relative position within each MHz (e.g., 1065.599 MHz and 1098.899 MHz). The middle sequence is labelled TF, for ``Triple Feature'', and consists of 5 signals that have the same frequency structure as the examples shown in Figure \ref{fig:TF} and are consistent with integer multiples of 133.33\,MHz. The bottom sequence is labelled SF, for ``Single Feature'', and consists of 7 signals that show the same frequency structure. All of the SF signals are spaced by multiples of 15\,MHz, except for the first pair, which is separated by 128.003\,MHz.}
    \label{fig:TF_SF_aliases.jpg}
\end{figure}

\begin{figure}
    \centering
    \includegraphics[width=\textwidth]{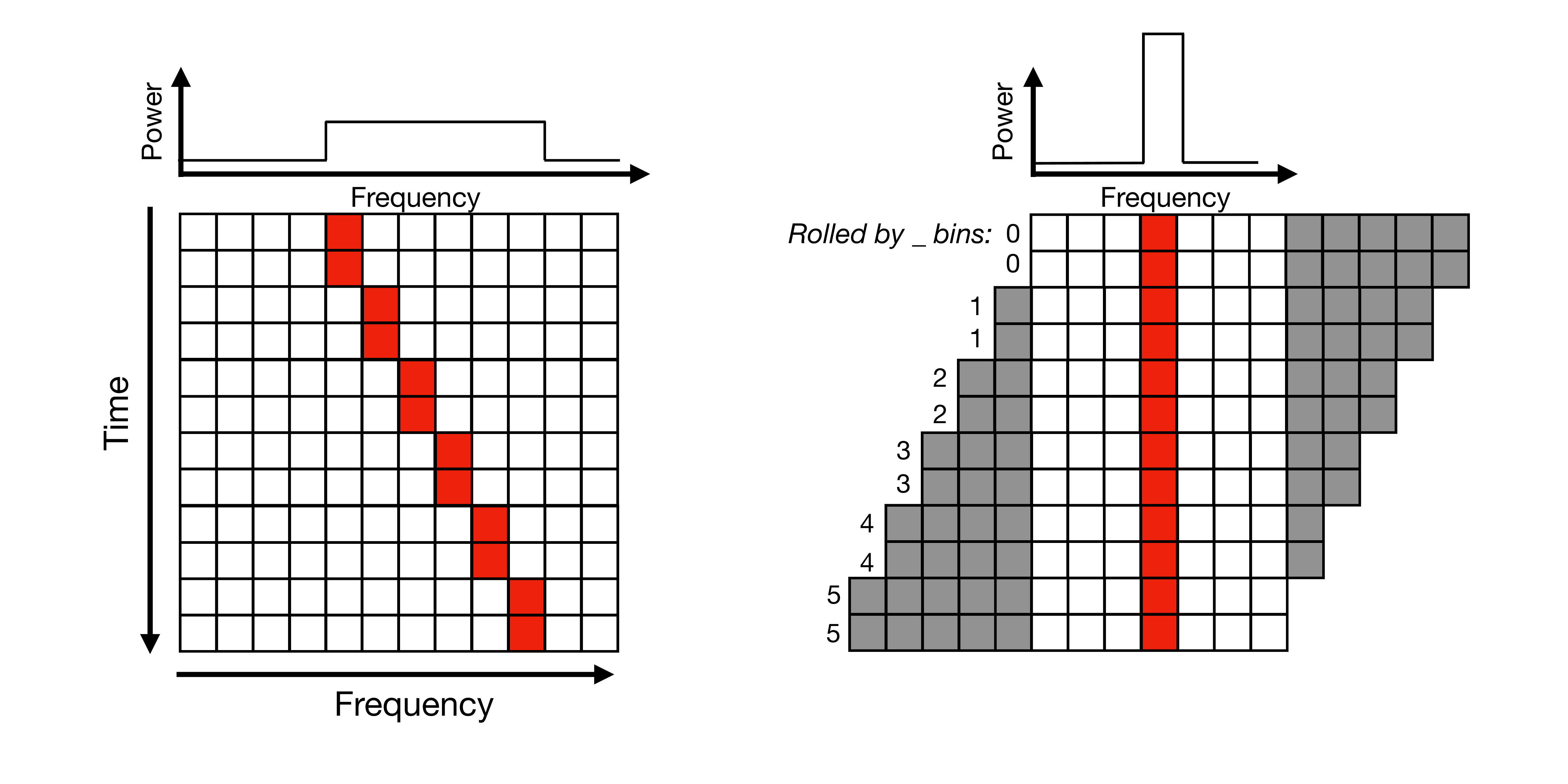}
    \caption{A cartoon illustrating the process of de-drifting a waterfall plot by ``rolling'' the rows. The graphic on the left shows a simplified waterfall plot with a drifting, unresolved signal in red. The power from this signal is spread over the majority of the frequency range when integrated over time, as shown in the power spectrum above the grid. However, as shown in the right graphic, when each row is shifted to align the signal in a single frequency bin via discretely shifting each pixel based on its width and the desired drift rate, the power in a single bin increases greatly, indicating a good fit to the actual drift rate of the signal. The frequencies at the edges cannot be used with this method (as indicated in grey in the cartoon); with the \texttt{numpy.roll} method \cite{harris2020array}, each row would actually wrap such that the gray boxes would be consolidated on the right of the figure. Regardless, less than 1\% of each waterfall plot falls in these edge pixels for the actual data used in this analysis.}
    \label{fig:dedrift_roll}
\end{figure}

\begin{figure}[ht]
    \centering
    \includegraphics[width=0.5\textwidth]{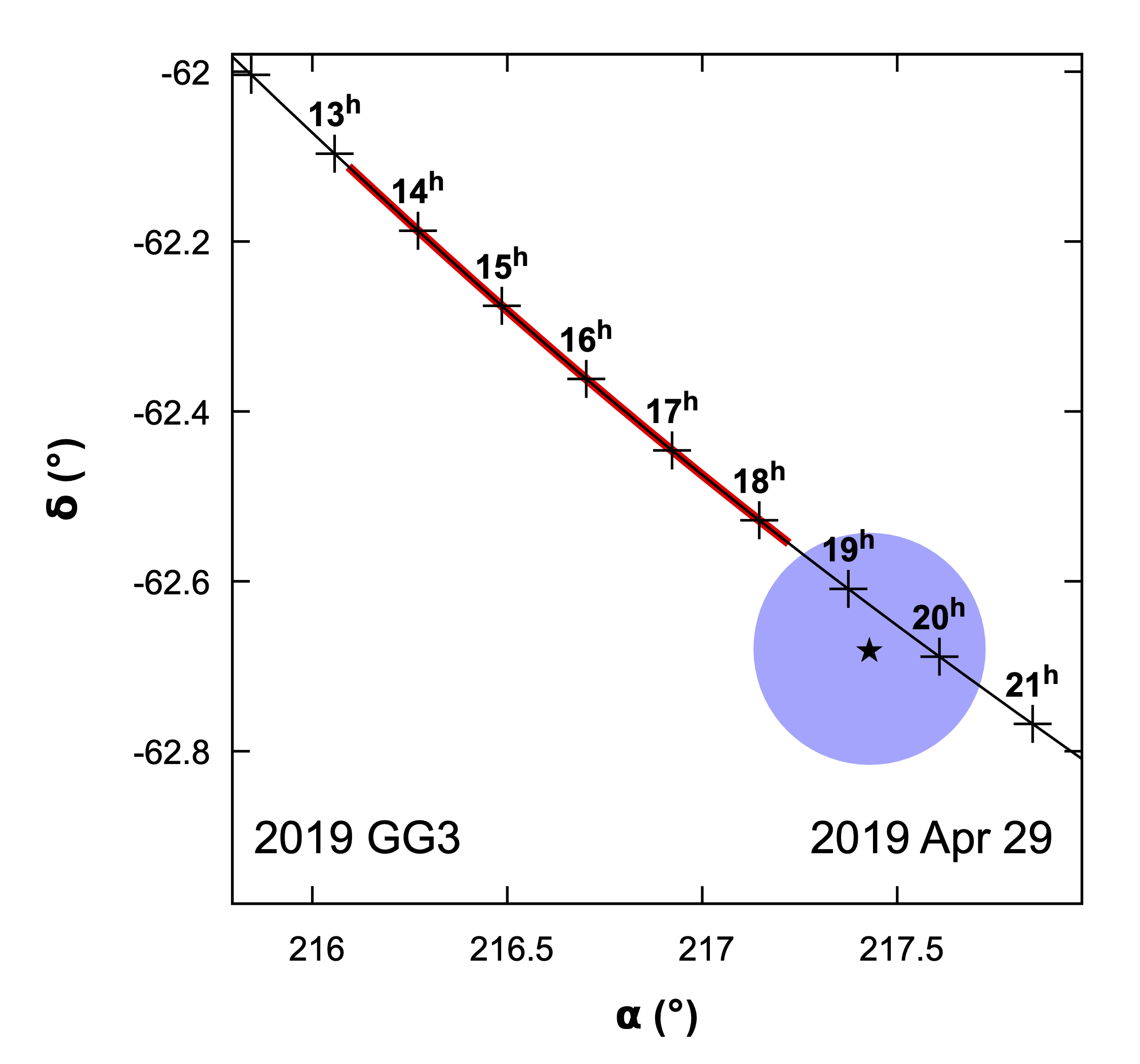}
    \caption{The track of 2019\,GG$_3$ relative to \ac{ProxCen} (star) on 2019 April 29 as it appeared from Parkes.  The blue disk roughly indicates the Murriyang 64-m beam size at 982\,MHz. Red highlighting indicates the duration of the blc1 observations. \label{fig:AsteroidTrack}}
\end{figure}

\begin{figure}[ht]
    \centering
    \includegraphics[width=0.5\textwidth]{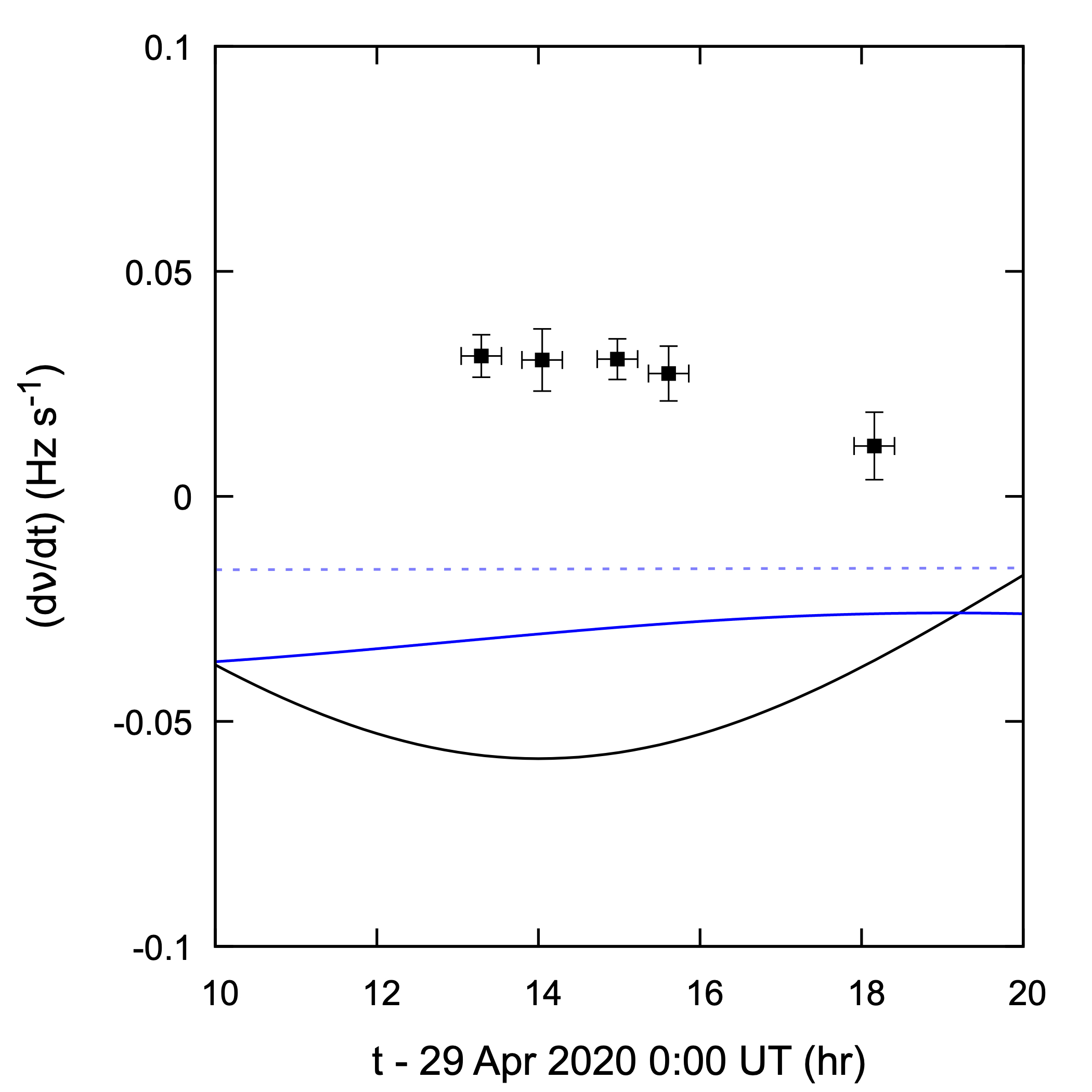}
    \caption{Drift rate for 2019\,GG$_3$ compared to that measured for blc1 (black points).  The pale blue dashed line shows the frequency drift due to 2019\,GG$_3$'s orbital motion alone (ignoring Earth's rotation), as observed from the Earth's center.  The drift rate of 2019\,GG$_3$ itself as viewed from Parkes (black line), including Earth's rotation and orbital motions, is compared with the best-fit reflector model from an Earthbound transmitter (blue line). Neither a transmitter on GG$_3$ itself nor a reflection off of GG$_3$ from an Earthbound transmitter can explain the observed drift of blc1.
    \label{fig:AsteroidDrift}}
\end{figure}

\begin{figure}[ht]
    \centering
    \includegraphics[width=0.8\textwidth]{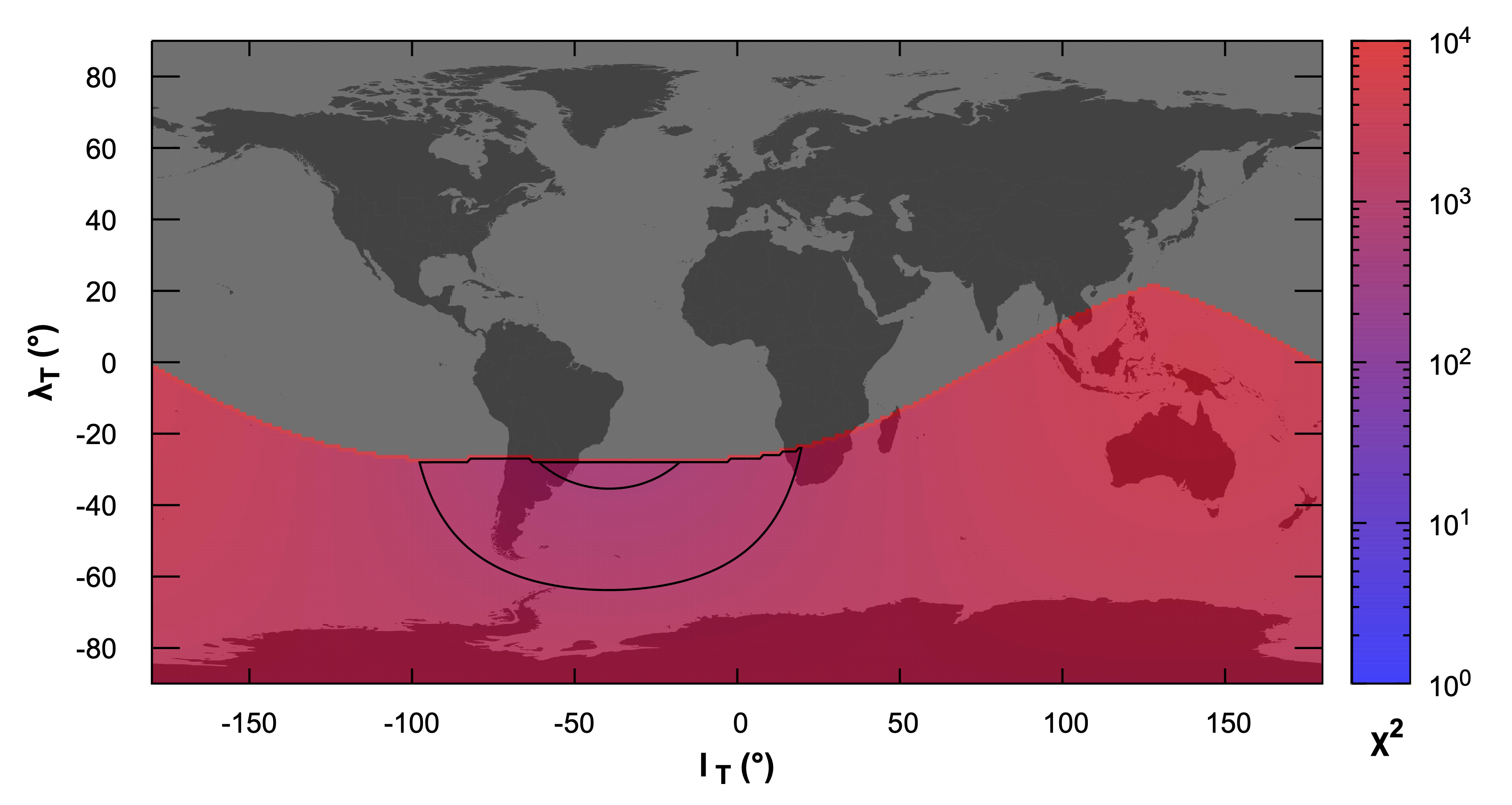}
    \caption{A map of the Earth showing the $\chi^2$ when fitting each possible transmitter location with 2019\,GG$_3$ as the reflector (predicted) to the drifts seen in blc1 (measured). Contours are for $\chi^2$ of $600$ and $1000$. No Earthbound transmitter location can explain the drift rate of blc1 in reflection off of GG$_3$.  (World map credit: Tom Patterson \textit{et al.}, via Wikimedia, \url{https://commons.wikimedia.org/wiki/File:Blank_Map_Equirectangular_states.svg}) \label{fig:EarthGrid-Asteroid}}
\end{figure}

\begin{figure}[ht]
    \centering
    \includegraphics[width=\textwidth]{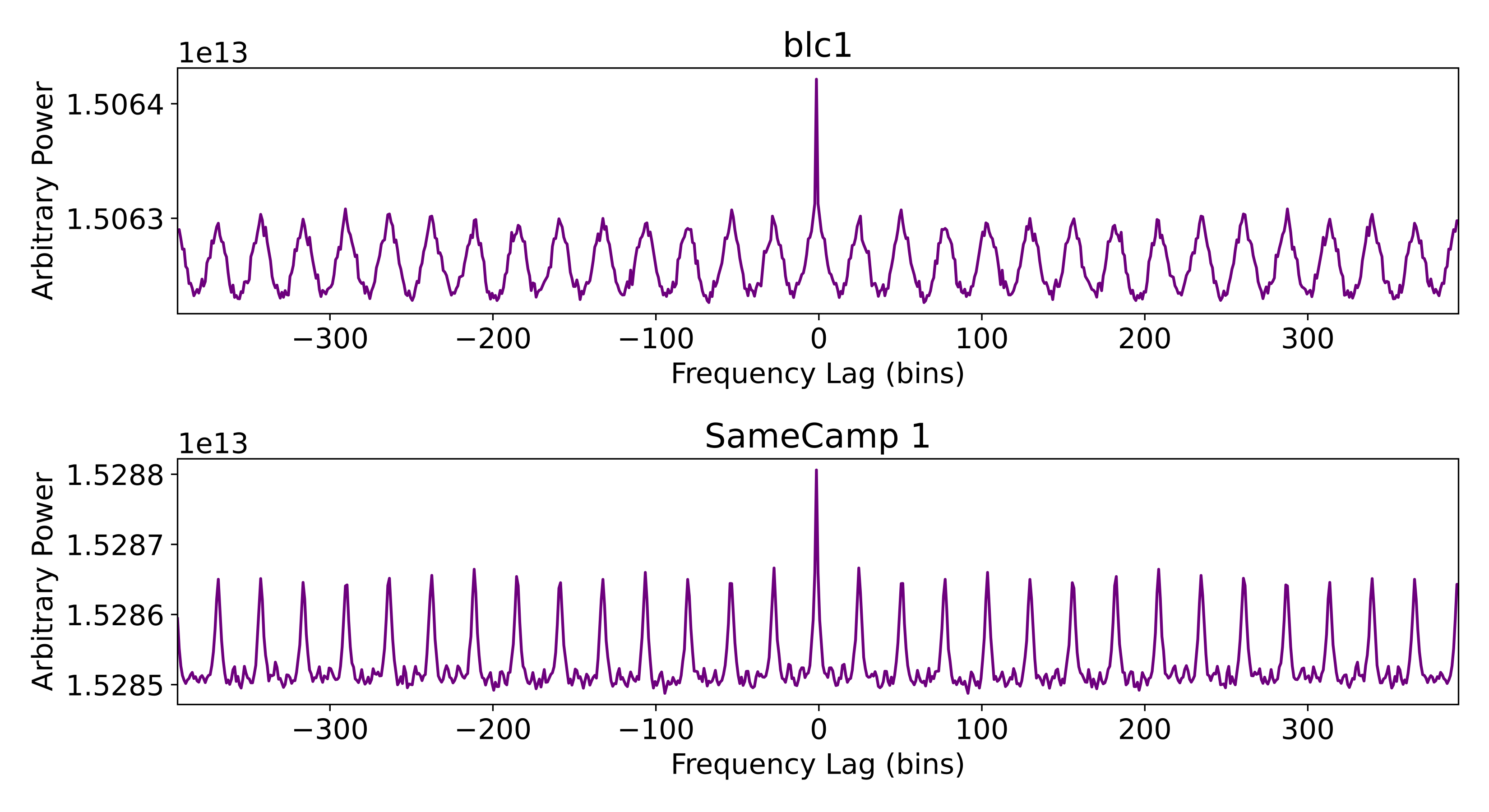}
    \caption{Auto-correlation functions of an observation of blc1 (top) and an observation of a potential redetection (bottom) in frequency, taken at similar UTCs one day apart. Both observations have signs of a periodic structure at a frequency lag of about 25 channels ($\sim 100$\,Hz, which corresponds to the comb-like structure visible in Figure 3.} 
    \label{fig:acfs}
\end{figure}

\begin{figure}[ht]
    \centering
    \includegraphics[width=\textwidth]{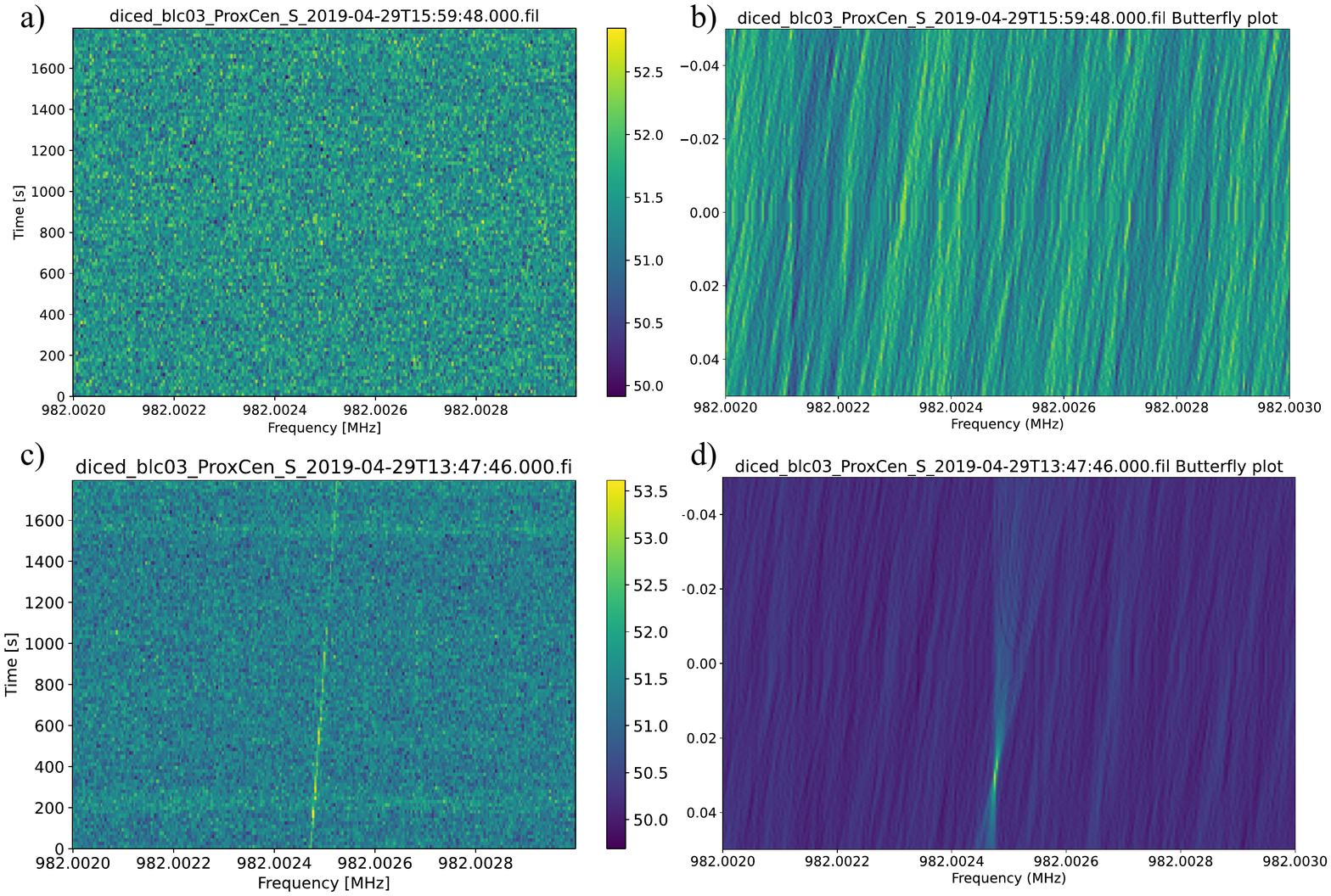}
    \caption{Two example waterfall plots a) and c) and two example butterfly plots b) and d). The first row contains plots from one of the 30-minute ProxCen observations without a detection from 29 April. The waterfall plot a) looks like noise at a glance, but the butterfly plot b) reveals traces of the \ac{RFI} comb mentioned in Section 2.1. Neither plot shows any trace of blc1. The second row contains plots from the second 30-minute detection of blc1. Note the feature offset at a drift rate of about 0.03\,Hz/s in d): the butterfly plot highlights linear features from the waterfall plots in a manner conducive to visual inspection.} 
    \label{fig:waterfall_butterfly}
\end{figure}

\begin{acronym}
\acro{SETI}{the Search for Extraterrestrial Intelligence}
\acro{SNR}{signal-to-noise ratio}
\acro{RFI}{Radio Frequency Interference}
\acro{BL}{Breakthrough Listen}
\acro{ProxCen}{Proxima Centauri}
\acro{UWL}{Ultra-Wide  bandwidth,  Low-frequency receiver}
\acro{NEO}{Near-Earth Object}
\acro{ETI}{Extraterrestrial Intelligence}
\acro{RA}{right ascension}
\acro{Dec}{declination}
\end{acronym}